\documentclass[12pt,UTF8]{article}
\usepackage{bm}
\usepackage{verbatim}
\usepackage{amssymb,amsmath,multicol,titlesec}

\usepackage{natbib}
\setcitestyle{authoryear,round,semicolon,aysep={,},yysep={},notesep={ }}
\usepackage[hmargin = 1in, vmargin = 1in,lmargin=1in, rmargin=1in]{geometry}

\usepackage[parfill]{parskip}
\usepackage{graphicx}
\usepackage{algorithm}
\usepackage{algorithmic}
\usepackage{subfig}
\usepackage{multirow}
\usepackage{mfirstuc}
\usepackage{xcolor}
\usepackage{longtable}
\usepackage{array}
\usepackage{makecell}
\usepackage[colorlinks,linkcolor=black]{hyperref}
\usepackage[labelfont={small},textfont={it,small}]{caption}
\usepackage{titlesec}
\usepackage{booktabs}
\usepackage{diagbox}
\usepackage{mathrsfs}
\usepackage{indentfirst}
\usepackage{makecell}
 \usepackage{hyperref}
 \hypersetup{hypertex=true,
            colorlinks=true,
            linkcolor=green,
            anchorcolor=red,
            menucolor=red,
            citecolor=green}
\titleformat{\section}[hang]{\fontsize{18pt}{1em}\selectfont \bfseries}{\thesection. }{0pt}{}
\titleformat{\subsection}[hang]{\fontsize{14pt}{1em}\selectfont \bfseries}{\thesubsection}{5pt}{}
\titleformat{\subsubsection}[hang]{\fontsize{12pt}{1em}\selectfont \bfseries}{\thesubsubsection}{5pt}{}
\titlespacing{\section}{0pt}{10pt}{10pt}
\titlespacing{\subsection}{0pt}{10pt}{-5pt}
\titlespacing{\subsubsection}{0pt}{8pt}{4pt}
\numberwithin{equation}{section}

\begin{document}

\begin{center}
\setlength{\baselineskip}{25pt}
\vspace{0.2in}
\noindent{\fontsize{18pt}{1em}\selectfont \textbf{A Two-Sample Robust  Bayesian Mendelian Randomization Method Accounting for  Linkage Disequilibrium and Idiosyncratic Pleiotropy with Applications to the COVID-19 Outcome  }}\\[12pt]

{\fontsize{14pt}{1.2em}\selectfont
Anqi Wang, Wei Liu, Zhonghua Liu\textsuperscript{*}
\\[10pt]
Department of Statistics and Actuarial Science, University of Hong Kong, Hong Kong SAR, China
\\[10pt]
\textsuperscript{*} To whom correspondence should be addressed: zhhliu@hku.hk \\[10pt]
\today
\\[0.6in]

}
\end{center}
\setlength{\baselineskip}{20pt}	

{\fontsize{18pt}{1em}\selectfont 
\centerline{\textbf{ABSTRACT}}
}
\vspace{10pt}
Mendelian randomization (MR) is a statistical method exploiting genetic variants as instrumental variables  to estimate the causal effect of modifiable risk factors on an outcome of interest. Despite wide uses of various popular two-sample MR methods based on genome-wide association study  summary level data, however, those methods could suffer from potential power loss or/and biased inference when the chosen genetic variants are in linkage disequilibrium (LD), and also have relatively large direct effects on the outcome whose distribution might be heavy-tailed which is commonly referred to as the idiosyncratic pleiotropy phenomenon. To resolve those two issues, we propose a novel Robust Bayesian Mendelian Randomization (RBMR) model that  uses the more robust multivariate generalized $t$-distribution \citep{arellano1995some} to model such direct effects in a probabilistic model framework which can also incorporate the LD structure explicitly.  The generalized $t$-distribution can be  represented  as a Gaussian scaled mixture so that our model parameters can be estimated by the expectation maximization(EM)-type algorithms. We compute the standard errors by calibrating the evidence lower bound  using the  likelihood ratio test.  Through extensive simulation studies, we show that our RBMR has robust performance compared to other competing methods.  We further apply our RBMR method to  two  benchmark data sets and find that RBMR has smaller bias and standard errors.  Using our proposed  RBMR method, we find that coronary artery disease (CAD) is associated with increased risk of critically ill coronavirus disease 2019 (COVID-19).  We also develop a user-friendly R package {\it RBMR} (\url{https://github.com/AnqiWang2021/RBMR}) for public use.\\
\noindent \textbf{Key\ Words: } COVID-19 outcome; Mendelian randomization; Idiosyncratic pleiotropy; Linkage disequilibrium; Multivariate generalized $t$-distribution; EM-type algorithm

\setlength{\parindent}{2em}
\newpage

\section{Introduction}

\noindent Mendelian randomization (MR) is a useful statistical method that leverages genetic variants as instrumental variables (IVs) for assessing the causal effect of a modifiable risk factor  on a health outcome of interest even in the presence of unmeasured confounding factors \citep{ebrahim2008mendelian,lawlor2008mendelian, evans2015mendelian}.  Because of the inborn nature of genetic variants, the associations between genetic variants and phenotypes after adjusting for possible population stratification will not be confounded by the environmental factors, socio-economic status and life styles after birth. 
Genome-wide association studies (GWAS) have identified tens of thousands of common genetic variants associated with thousands of complex traits and diseases \citep{macarthur2017new}. Those  GWAS summary level data  contain rich information about genotype-phenotype associations (\url{https://www.ebi.ac.uk/gwas/}), and thus provide us valuable resources for MR studies. Therefore, we have seen a boost of two-sample MR method developments and applications based on GWAS summary statistics recently due to the increasing availability of candidate genetic variant IVs for thousands of phenotypes.  \citep{burgess2013mendelian,bowden2015mendelian,pickrell2016detection}.
In particular, a genetic variant serving as a valid IV must satisfy the following three core assumptions \citep{martens2006instrumental,lawlor2008mendelian}:
\begin{enumerate}
\item[1.] \textbf{Relevance:} The genetic variant must be associated (not necessarily causally) with the exposure;
\item[2.] \textbf{Effective Random Assignment:} The genetic variant must be independent of any (measured or unmeasured) confounders of the exposure-outcome relationship;
\item[3.] \textbf{Exclusion Restriction:} The genetic variant must affect the outcome only through the exposure, that is, the genetic variant must have no direct effect on the outcome not mediated by the exposure. 
\end{enumerate}

When these three core IV assumptions hold, the inverse variance weighted (IVW)  \citep{ehret2011genetic} method can be simply used to obtain unbiased causal effect estimate of the exposure on the outcome. However, among those three core assumptions, only the IV relevance assumption can be empirically tested, for example, by checking the empirical association strength between the candidate IV and the exposure using the GWAS catalog (\url{https://www.ebi.ac.uk/gwas/}). The association between the IV and the exposure must be strong enough (the IV explains a large amount of the variation of the exposure variable) to ensure unbiased causal effect estimate. The problem of weak IVs has been studied previously in the econometric literature \citep{bound1995problems,hansen2008estimation}. In MR settings,  the method that uses genetic score by combining multiple weak IVs together to increase the IV-exposure association strength to reduce weak IV bias has also been proposed \citep{evans2013mining}. Unfortunately, the other two IV core assumptions cannot be empirically tested and might be violated in practice.   Violation of the exclusion restriction assumption can occur when the genetic variant indeed has a non-null direct effect on the outcome not mediated by the exposure, referred to as systematic pleiotropy \citep{solovieff2013pleiotropy, verbanck2018detection, zhao2020statistical}.  However, very often, genetic variants might have relatively large direct effects whose distribution exhibits a heavy-tailed pattern, a phenomenon referred to as the idiosyncratic pleiotropy in this paper. For example, there exists idiosyncratic pleiotropy when estimating the causal effect of low-density lipoprotein (LDL) cholesterol on the risk of Alzheimer's disease. In Section \ref{real}, we will describe more details about this real data example. \\
\indent To address those possible violations of the IV core assumptions and potential risk, many efforts have been made recently. The MR-Egger regression method introduced an intercept term to capture the presence of unbalanced systematic pleiotropy under the Instrument Strength Independent of Direct Effect (InSIDE)  assumption \citep{bowden2015mendelian}. However, MR-Egger would be biased when there exists idiosyncratic pleiotropy. \cite{zhu2018causal} proposed the GSMR method that  removes suspected genetic variants with relatively large direct effects and also takes the LD structure into account by using the generalized least squares approach. However, removal of a large number of relatively large direct effects might lead to efficiency loss.  \citet{zhao2020statistical} proposed MR-RAPS to improve statistical power for causal inference and limit the influence of relatively large direct effects by using the adjusted profile likelihood and robust loss functions assuming that those SNP IVs are independent. However, this independent IV assumption  might not hold in practice because SNPs within proximity tend to be correlated.  \citet{cheng2020mr} proposed a  two-sample MR method  named MR-LDP that built a Bayesian probabilistic model accounting for systematic pleiotropy and LD structures among SNP IVs.  One drawback of the MR-LDP  method is that it cannot handle relatively large direct effects well. \\
\indent To overcome the limitations of  those aforementioned methods, we propose a more robust method named `Robust Bayesian Mendelian Randomization (RBMR)' accounting for LD, systematic and idiosyncratic pleiotropy simultaneously in a unified framework. Specifically, to account for LD, we first estimate the LD correlation matrix of SNP IVs and then explicitly include it in the model likelihood. To account for idiosyncratic pleiotropy, we propose to model the direct effects using the more robust multivariate generalized $t$-distribution \citep{arellano1995some, frahm2004generalized} which will be shown to have improved performance than using the Gaussian distribution when the idiosyncratic pleiotropy is present.  Moreover, this more robust distribution can be represented as a Gaussian scaled mixture to facilitate  model parameter estimation using the parameter expanded variational Bayesian expectation maximization algorithm (PX-VBEM) \citep{yang2020comm} which combines the VB-EM \citep{beal2003variational} and the PX-EM \citep{liu1998parameter} together.  We further calculate the standard error by calibrating the evidence lower bound (ELBO) according to a nice property of the likelihood ratio test (LRT).  Both extensive simulation studies  in Section \ref{simulations}  and  analysis of two real benchmark data sets  in Section \ref{real} show that our proposed  RBMR method outperforms competitors. The real data analysis results show that coronary artery disease (CAD) is associated with increased risk of critically ill COVID-19 outcome.
\section{Methods}
\subsection{The Linear Structural Model}
\noindent Suppose that we have $J$ possibly correlated genetic variants (for example, single-nucleotide polymorphisms, or SNPs ) $G_{j}, j=1,2,\dots, J$, the exposure variable $X$, the outcome variable $Y$ of interest and  unknown confounding factors $U$.  Let $\delta_{X}$ and $\delta_{Y}$ denote the effects of confounders $U$ on exposure $X$ and outcome $Y$ respectively. The coefficients $\gamma_{j}\ (j=1,2,\dots, J)$ denote the SNP-exposure true effects. Suppose that all the IVs are valid, then the exposure can be represented as a  linear structural function of the SNPs,  confounders and an independent random noise term $e_{X}$. The outcome can be represented as a linear structural function of the exposure, confounders and the independent random noise term $e_{Y}$. The true effect size of the exposure on the outcome is denoted as $\beta_0$. Then, we have the following linear structural equation models \citep{bowden2015mendelian}: 
\begin{equation}\label{a}
X=\sum_{j=1}^{J} G_{j} \gamma_{j}+U \delta_{X}+e_{X}, \quad Y=\beta_{0} X+U\delta_{Y}+e_{Y}.
\end{equation}
Let $\Gamma_{j}\ (j=1,2,\dots, J)$ be the  true effects of SNPs on the outcome.  With valid IVs,  we have 
\begin{equation}\label{2.3}
\Gamma_{j}=\beta_{0} \gamma_{j}.
\end{equation}
To accommodate  possible violations of the exclusion restriction assumption,   we now consider the following modified linear structural functions \citep{bowden2015mendelian}:
\begin{equation}\label{b}
X=\sum_{j=1}^{J} G_{j} \gamma_{j}+U \delta_{X}+e_{X}, \quad Y=\sum_{j=1}^{J} G_{j} \alpha_{j}+\beta_{0} X+U\delta_{Y} \textcolor{blue}{+} e_{Y},
\end{equation}
where the coefficients $\alpha_j\ (j=1,2,\dots, J)$ represent the direct effects of the SNPs on the outcome. Then we have
\begin{equation}\label{m}
\Gamma_{j}=\beta_{0} \gamma_{j}+\alpha_{j}.
\end{equation}

So far, many existing MR methods assign the Gaussian distribution on each direct effect $\alpha_j$, that is $\bm{\alpha}\sim \mathcal{N}(\bm{0},\sigma_0^2\mathbf{I_J})$ \citep{zhao2020statistical,cheng2020mr,zhao2020bayesian},  where $\bm{\alpha}=[\alpha_1,\dots,\alpha_J]^{\mathrm{T}}$ is a $J$-dimensional vector of direct effects. However, real genetic data might contain some relatively large direct effects whose distribution can be heavy-tailed,  and thus the Gaussian distribution might not be a good fit. Therefore, we propose  to assign the multivariate generalized $t$-distribution on $\bm{\alpha}$ \citep{arellano1995some, kotz2004multivariate}, which is  a robust alternative to the Gaussian distribution \citep{frahm2004generalized}. 
%
 \subsection{The Robust Bayesian MR Model}

\noindent Let $\left\{\widehat{\gamma}_{j}, \widehat{\sigma}_{X_{j}}^{2}\right\}_{j=1, \dots, J}$ and $\left\{\widehat{\Gamma}_{j}, \widehat{\sigma}_{Y_{j}}^{2}\right\}_{j=1, \dots, J}$ be the GWAS summary statistics for the exposure and the outcome respectively, where $\left\{\widehat{\sigma}_{X_{j}}^{2}, \widehat{\sigma}_{Y_{j}}^{2}\right\}$ are the corresponding estimated standard errors. Many existing MR methods assume that IVs are independent from each other \citep{ehret2011genetic, bowden2015mendelian, zhao2020statistical}, and the uncorrelated SNPs can be chosen by using a tool called LD clumping \citep{hemani2016mr,purcell2007plink}, which might remove many SNP IVs and thus cause efficiency loss. 
To include more SNP IVs even if they are in  LD, we need to account for the LD structure explicitly. To achieve this goal, we use a reference panel sample to assist with reconstructing LD matrix, such as the 1000 Genome Project Phase 1 ($N$=379) \citep{10002012integrated}. We first apply the LDetect method to partition the whole genome into $Q$ blocks \citep{berisa2016approximately} and then estimate the LD matrix $\bm{\Theta}$ using the estimator $\widehat{\bm{\Theta}}^{(k)}(k=1,2,\dots,Q)$  first proposed by \citet{rothman2012positive}. 
Then, the distributions of $\widehat{\bm{\gamma}}$ and $\widehat{\bm{\Gamma}}$ are given by
\begin{equation}\label{ccc}
\widehat{\bm{\gamma}} | \bm{\gamma}, \widehat{\bm{\Theta}}, \widehat{\bm{\sigma}}_{\bm{X}} \sim \mathcal{N}\left(\widehat{\bm{\sigma}}_{\bm{X}} \widehat{\bm{\Theta}} \widehat{\bm{\sigma}}_{\bm{X}}^{-1} \bm{\gamma}, \widehat{\bm{\sigma}}_{\bm{X}} \widehat{\bm{\Theta}} \widehat{\bm{\sigma}}_{\bm{X}}\right),
\end{equation}
\begin{equation}\label{d}
\widehat{\bm{\Gamma}} | \bm{\Gamma}, \widehat{\bm{\Theta}}, \widehat{\bm{\sigma}}_{\bm{Y}} \sim \mathcal{N}\left(\widehat{\bm{\sigma}}_{\bm{Y}} \widehat{\bm{\Theta}} \widehat{\bm{\sigma}}_{\bm{Y}}^{-1} \bm{\Gamma}, \widehat{\bm{\sigma}}_{\bm{Y}} \widehat{\bm{\Theta}} \widehat{\bm{\sigma}}_{\bm{Y}}\right),
\end{equation}
where $\widehat{\bm{\sigma}}_{\bm{X}}=\operatorname{diag}\left(\left[\widehat{\sigma}_{X_{1}}, \dots, \widehat{\sigma}_{X_{J}}\right]\right)$ and $\widehat{\bm{\sigma}}_{\bm{Y}}=\operatorname{diag}\left(\left[\widehat{\sigma}_{Y_{1}}, \dots, \widehat{\sigma}_{Y_{J}}\right]\right)$ are both diagonal matrices  \citep{zhu2017bayesian}. \\
\indent To account for the presence of idiosyncratic pleiotropy,  we propose to model the direct effects $\bm{\alpha }$ using the more robust multivariate generalized $t$-distribution \citep{arellano1995some, kotz2004multivariate,ala2016gaussian} whose density function is given by 
\begin{equation}\label{gau}
\begin{split}
t_{J}(\bm{\alpha}|\mathbf{\Sigma}, \alpha_w,\beta_w)&=\frac{f(\alpha_w+J / 2)}{|\mathbf{\Sigma}|^{1 / 2} f(\alpha_w )(2\pi \beta_w)^{J / 2}}\left[1+\frac{1}{2\beta_w}(\bm{\alpha}^{\mathrm{T}}\mathbf{\Sigma}^{-1}\bm{\alpha}) \right]^{-(\alpha_w+J / 2)}\\
&=\int \mathcal{N}(\bm{\alpha} | \boldsymbol{0}, \boldsymbol{\Sigma} / w) \mathcal{G}(w | \alpha_w, \beta_w) \mathrm{d} w, 
\end{split}
\end{equation}
where $\mathcal{N}(\bm{\alpha} | \boldsymbol{0}, \boldsymbol{\Sigma} / w)$ denotes the $J$-dimensional Gaussian distribution with mean $\boldsymbol{0}$ and covariance $\boldsymbol{\Sigma}/w$, $\mathbf{\Sigma}=\sigma_0^2\mathbf{I_J}$ is a $J\times J$ diagonal matrix, and $\mathcal{G}(w |\alpha_w, \beta_w)$ is the Gamma distribution of a univariate positive variable $w$ referred to as a weight variable
\begin{equation}\label{G}
\mathcal{G}(w| \alpha_w, \beta_w)=\frac{{\beta_w}^{\alpha_w}}{f({\alpha_w})}w^{\alpha_w-1}e^{-\beta_w w},
\end{equation}
where $f$ denotes the Gamma function. When $\alpha_w=\beta_w=\nu/2$ in equation (\ref{G}), the distribution in equation (\ref{gau}) reduces to a multivariate $t$-distribution, where $\nu$ is the degree of freedom. Gaussian scaled mixture representation enables the use of EM-type algorithms for statistical inference, such as the PX-VBEM \citep{yang2020comm} described in Section \ref{algorithm}. \\
\indent Then we denote the distribution of the latent variable $\bm{\gamma}$ as
\begin{equation}\label{mm}
\bm{\gamma}|\bm{\sigma}^{2} \sim \mathcal{N} \left(\mathbf{0}, \bm{\sigma}^{2} \right),
\end{equation}
where $\bm{\sigma}^2=\sigma^2\mathbf{I}_{\mathbf{J}}$ is a $J\times J$ diagonal matrix. By assuming that $\bm{\gamma}$, $\bm{\alpha}$ and $w$ are latent variables, the complete data likelihood can be written as
\begin{equation}\label{ee}
\begin{split}
\operatorname{Pr}\left(\widehat{\bm{\Gamma}}, \widehat{\bm{\gamma}}, \boldsymbol{\alpha}, \bm{\gamma},w | \widehat{\bm{\sigma}}_{\bm{X}}, \widehat{\bm{\sigma}}_{\bm{Y}}, \widehat{\bm{\Theta}} ; \boldsymbol{\theta}, \boldsymbol{h}\right)=&\mathcal{N}\left(\widehat{\bm{\Gamma}} | \widehat{\bm{\sigma}}_{\bm{Y}} \widehat{\bm{\Theta}} \widehat{\bm{\sigma}}_{\bm{Y}}^{-1}\left(\beta_{0} \boldsymbol{\gamma}+\boldsymbol{\alpha}\right), \widehat{\bm{\sigma}}_{\bm{Y}} \widehat{\bm{\Theta}} \widehat{\bm{\sigma}}_{\bm{Y}}\right)\mathcal{N}\left(\mathbf{0}, {\sigma}^{2}\mathbf{I}_{\mathbf{J}}\right) \\
\times &\mathcal{N}\left(\widehat{\boldsymbol{\gamma}} |  \widehat{\bm{\sigma}}_{\bm{X}} \widehat{\bm{\Theta}} \widehat{\bm{\sigma}}_{\bm{X}}^{-1} \boldsymbol{\gamma}, \widehat{\bm{\sigma}}_{\bm{X}} \widehat{\bm{\Theta}} \widehat{\bm{\sigma}}_{\bm{X}}\right)  \mathcal{N}\left(\bm{\alpha}|\mathbf{0}, \sigma_{0}^{2}\mathbf{I}_{\mathbf{J}}/w\right) \mathcal{G}\left(w|\alpha_w,\beta_w\right).
\end{split}
\end{equation} 

\subsection{Estimation and Inference}\label{algorithm}
\noindent The standard expectation-maximization (EM) algorithm \citep{dempster1977maximum} is a popular choice for finding the maximum likelihood estimate  in the presence of missing (latent) variables.  However, one difficulty for implementing the EM algorithm is to calculate the marginal likelihood function which might involve difficult integration with respect to the distributions of the latent variables. In addition, the original EM algorithm might be slow \citep{liu1998parameter}.
 To address these numerical issues, we utilize a parameter expanded variational Bayesian expectation-maximization algorithm, namely, PX-VBEM \citep{yang2020comm}, by replacing the EM algorithm in VB-EM \citep{beal2003variational} with PX-EM algorithm  \citep{liu1998parameter} to accelerate the speed of convergence. To start with, for the purpose of applying the PX-EM algorithm,  the distribution of $\widehat{\bm{\gamma}}$ in equation (\ref{ccc}) can be rewritten as follows:
\begin{equation}
\widehat{\bm{\gamma}} | \bm{\gamma}, \widehat{\bm{\Theta}}, \widehat{\bm{\sigma}}_{\bm{X}} \sim \mathcal{N}\left(\zeta\widehat{\bm{\sigma}}_{\bm{X}} \widehat{\bm{\Theta}} \widehat{\bm{\sigma}}_{\bm{X}}^{-1} \bm{\gamma}, \widehat{\bm{\sigma}}_{\bm{X}} \widehat{\bm{\Theta}} \widehat{\bm{\sigma}}_{\bm{X}}\right).
\end{equation}
We also rewrite the complete data likelihood in equation (\ref{ee}) as: 
\begin{equation}
\begin{split}
\operatorname{Pr}\left(\widehat{\bm{\Gamma}}, \widehat{\bm{\gamma}}, \boldsymbol{\alpha}, \bm{\gamma},w | \widehat{\bm{\sigma}}_{\bm{X}}, \widehat{\bm{\sigma}}_{\bm{Y}}, \widehat{\bm{\Theta}} ; \boldsymbol{\theta}, \boldsymbol{h}\right)=&\mathcal{N}\left(\widehat{\bm{\Gamma}} | \widehat{\bm{\sigma}}_{\bm{Y}} \widehat{\bm{\Theta}} \widehat{\bm{\sigma}}_{\bm{Y}}^{-1}\left(\beta_{0} \boldsymbol{\gamma}+\boldsymbol{\alpha}\right), \widehat{\bm{\sigma}}_{\bm{Y}} \widehat{\bm{\Theta}} \widehat{\bm{\sigma}}_{\bm{Y}}\right)\mathcal{N}\left(\mathbf{0}, {\sigma}^{2}\mathbf{I}_{\mathbf{J}}\right) \\
\times &\mathcal{N}\left(\widehat{\boldsymbol{\gamma}} |  \zeta\widehat{\bm{\sigma}}_{\bm{X}} \widehat{\bm{\Theta}} \widehat{\bm{\sigma}}_{\bm{X}}^{-1} \boldsymbol{\gamma}, \widehat{\bm{\sigma}}_{\bm{X}} \widehat{\bm{\Theta}} \widehat{\bm{\sigma}}_{\bm{X}}\right)  \mathcal{N}\left(\mathbf{0}, \sigma_{0}^{2}\mathbf{I}_{\mathbf{J}}/w\right) \mathcal{G}\left(w | \alpha_w, \beta_w\right) ,
\end{split}
\end{equation} 
where the expanded model parameters for RBMR are $\boldsymbol{\theta} \stackrel{\text { def }}{=} \left\{\beta_{0}, \sigma_{0}^{2}, \sigma^{2},\zeta \right\}$. Let $q(\bm{\gamma},\bm{\alpha},w)$ be a variational posterior distribution.  The logarithm of the marginal likelihood  can be decomposed into two parts,
\begin{equation}
\begin{split}
&\log \operatorname{Pr}\left(\widehat{\boldsymbol{\gamma}}, \widehat{\Gamma} | \widehat{\bm{\sigma}}_{\bm{X}}, \widehat{\bm{\sigma}}_{\bm{Y}}, \widehat{\bm{\Theta}} ; \boldsymbol{\theta}, \boldsymbol{h}\right)\\
&=\mathbb{E}_{q(\bm{\gamma},\bm{\alpha},w)}\left[\log \operatorname{Pr}\left(\widehat{\bm{\gamma}}, \widehat{\bm{\Gamma}} |\widehat{\bm{\sigma}}_{\bm{X}}, \widehat{\bm{\sigma}}_{\bm{Y}},  \widehat{\bm{\Theta}} ; \boldsymbol{\theta}, \boldsymbol{h}\right)\right]\\
&=\mathcal{L}(q)+\mathbb{K} \mathbb{L}(q \| p),
\end{split}
\end{equation}
where
\begin{equation}
\begin{split}
\mathcal{L}(q)&=\mathbb{E}_{q(\bm{\gamma},\bm{\alpha},w)}\left[\log \frac{\operatorname{Pr}\left(\widehat{\bm{\gamma}}, \widehat{\bm{\Gamma}}, \bm{\gamma},\bm{\alpha},w | \widehat{\bm{\sigma}}_{\bm{X}}, \widehat{\bm{\sigma}}_{\bm{Y}}, \widehat{\bm{\Theta}} ; \boldsymbol{\theta}, \boldsymbol{h}\right)}{q\left(\bm{\gamma},\bm{\alpha},w\right)}\right],\\
\mathbb{K} \mathbb{L}(q \| p)&=\mathbb{E}_{q(\bm{\gamma},\bm{\alpha},w)}\left[\log \frac{q\left(\bm{\gamma},\bm{\alpha},w\right)}{p\left(\bm{\gamma},\bm{\alpha},w|\widehat{\bm{\gamma}}, \widehat{\bm{\Gamma}}, \widehat{\bm{\sigma}}_{\bm{X}}, \widehat{\bm{\sigma}}_{\bm{Y}}, \widehat{\bm{\Theta}} ; \boldsymbol{\theta}, \boldsymbol{h}\right)}\right].
\end{split}
\end{equation}
Given that the $\mathcal{L}(q)$ is an evidence lower bound (ELBO) of the marginal log-likelihood, the non-negative Kullback-Leibler (KL) divergence $\mathbb{K} \mathbb{L}(q \| p)$ is equal to zero if and only if the variational posterior distribution is equal to the true posterior distribution. Minimizing the KL divergence is equivalent to maximizing ELBO. Before calculating the maximization of ELBO, due to the fact that latent variables are independent of each other, the decomposition form of the posterior distribution $q(\bm{\gamma},\bm{\alpha}, w)$ is obtained using the mean field assumption \citep{blei2017variational},
\begin{equation}
q(\bm{\gamma}, \boldsymbol{\alpha},w)=\prod_{j=1}^{J} q\left(\gamma_{j}\right) \prod_{j=1}^{J} q\left(\alpha_{j}\right)q(w).
\end{equation}

In the PX-VB-E step, the optimal variational posterior distributions for $\bm{\gamma}$, $\bm{\alpha}$ and $w$ can be written as: 
\begin{equation}\label{R1}
\begin{split}
q\left(\bm{\gamma}|\mu_{\gamma_j},\sigma_{\gamma_j}^2\right)=\prod_{j=1}^J \mathcal{N}\left(\mu_{\gamma_j},\sigma_{\gamma_j}^2\right)&,\ q\left(\bm{\alpha}|\mu_{\alpha_j},\sigma_{\alpha_j}^2\right)=\prod_{j=1}^J \mathcal{N}\left(\mu_{\alpha_j},\sigma_{\alpha_j}^2\right),\\
q\left(w|\widetilde{\alpha}_w,\widetilde{\beta}_w\right)&=\mathcal{G}\left(\widetilde{\alpha}_w,\widetilde{\beta}_w\right).
\end{split}
\end{equation}
 The updating equations for the parameters are given by
\begin{equation}\label{R2}
\begin{aligned}
-\frac{1}{2 \sigma_{\gamma_j}^{2}}&=-\frac{\beta_{0}^{2}}{2} \frac{\widehat{\bm{\Theta}}_{j j}}{\sigma_{Y_j}^{2}}-\frac{\zeta^{2} \widehat{\bm{\Theta}}_{j j}}{2 \sigma_{X_j}^{2}}-\frac{1}{2 \sigma^{2}},\\
\frac{\mu_{\gamma_j}}{\sigma_{\gamma_j}^{2}}&=\beta_{0} \frac{\widehat{\Gamma}_{j}}{\sigma_{Y_j}^{2}}-\frac{\beta_{0}^{2}}{\sigma_{Y_j}}\left(\sum_{j^{'} \neq j} \frac{\left[\gamma_{j^{'}}\right]\widehat{\bm{\Theta}}_{j j^{'}}}{\sigma_{Y_j}}\right)-\frac{\beta_{0}}{\sigma_{Y_ j}}\left(\sum_{j^{'}=1}^{J} \frac{\left[ \alpha_{j^{'}}\right] \widehat{\bm{\Theta}}_{j j^{'}}}{\sigma_{Y_j^{'}}}\right)+\frac{\zeta \widehat{\gamma}_{j}}{\sigma_{X_j}^{2}}-\frac{\zeta^{2}}{\sigma_{X_j}}\left(\sum_{j^{'} \neq j} \frac{\left[\gamma_{j^{'}}\right] \widehat{\bm{\Theta}}_{j j^{'}}}{\sigma_{X_j^{'}}}\right),\\
-\frac{1}{2 {\sigma}_{\alpha_j}^{2}}&=-\frac{1}{2} \frac{\widehat{\bm{\Theta}}_{j j}}{\sigma_{Y_j}^{2}}-\frac{\left[ w \right] }{2 \sigma_{0}^{2}},\\
\frac{{\mu}_{\alpha_j}}{{\sigma}_{\alpha_j}^{2}}&=\frac{\widehat{\Gamma}_{j}}{\sigma_{Y_j}^{2}}-\frac{\beta_{0}}{\sigma_{Y_j}} \sum_{j^{'}=1}^{J} \frac{\widehat{\bm{\Theta}}_{j j^{'}}\left[\gamma_{j^{'}}\right]}{\sigma_{Y_j^{'}}}-\frac{1}{\sigma_{Y_j}} \sum_{j^{'} \neq j} \frac{\left[\alpha_{j^{'}}\right] \widehat{\bm{\Theta}}_{j j^{'}}}{\sigma_{Y_j^{'}}},\\
\widetilde{\alpha}_w &= \alpha_w+ \frac{J}{2}, \\
\widetilde{\beta}_w &=\beta_w + \sum_{j=1}^J \frac{\left[ \alpha_j^2\right]}{\sigma_{0}^2}.
\end{aligned}
\end{equation}
where $\left[\gamma_{j^{'}}\right] \stackrel{\text { def }}{=} E_{q}\left(\gamma_{j^{'}}\right)$, $\left[\alpha_{j^{'}(j)}\right] \stackrel{\text { def }}{=} E_{q}\left(\alpha_{j^{'}(j)}\right)$ and $\left[ w \right] \stackrel{\text { def }}{=} E_{q}\left(w \right)$. 

In the PX-VB-M step, by setting the derivate of the ELBO to be  zero, the model parameters $\boldsymbol{\theta}$ can be  obtained as:
\begin{equation}\label{al5}
\begin{aligned}
\beta_{0}&=\left\{\boldsymbol{\mu}_{\bm{\gamma}}^{\mathrm{T}} \widehat{\bm{\sigma}}_{\bm{Y}}^{-1} \widehat{\bm{\Theta}} \widehat{\bm{\sigma}}_{\bm{Y}}^{-1} \boldsymbol{\mu}_{\bm{\gamma}}+\operatorname{Tr}\left(\widehat{\bm{\sigma}}_{\bm{Y}}^{-1} \widehat{\bm{\Theta}} \widehat{\bm{\sigma}}_{\bm{Y}}^{-1} \boldsymbol{\mathbf{S}}_{\bm{\gamma}}\right)\right\}^{-1}\left(\widehat{\bm{\Gamma}}^{\mathrm{T}} \widehat{\bm{\sigma}}_{\bm{Y}}^{-2} \boldsymbol{\mu}_{\bm{\gamma}}-\boldsymbol{\mu}_{\bm{\alpha}}^{\mathrm{T}} \widehat{\bm{\sigma}}_{\bm{Y}}^{-1} \widehat{\bm{\Theta}} \widehat{\bm{\sigma}}_{\bm{Y}}^{-1} \boldsymbol{\mu}_{\bm{\gamma}}\right),\\
\sigma^{2}&=\left\{\boldsymbol{\mu}_{\bm{\gamma}}^{\mathrm{T}} \boldsymbol{\mu}_{\bm{\gamma}}+\operatorname{Tr}\left(\boldsymbol{\mathbf{S}}_{\bm{\gamma}}\right)\right\} /J ,\\
\sigma_{0}^{2}&=\left\{\widetilde{\alpha}_w \left(\boldsymbol{\mu}_{\alpha}^{\mathrm{T}} \boldsymbol{\mu}_{\alpha}+\operatorname{Tr}\left(\boldsymbol{\mathbf{S}}_{\boldsymbol{\alpha}}\right)\right)\right\} /J\widetilde{\beta}_w ,\\
\zeta&=\left\{\mu_{\bm{\gamma}}^{\mathrm{T}} \widehat{\bm{\sigma}}_{\bm{X}}^{-1} \widehat{\bm{\Theta}} \widehat{\bm{\sigma}}_{\bm{X}}^{-1} \mu_{\bm{\gamma}}+\operatorname{Tr}\left(\widehat{\bm{\sigma}}_{\bm{X}}^{-1} \widehat{\bm{\Theta}} \widehat{\bm{\sigma}}_{\bm{X}}^{-1} \boldsymbol{\mathbf{S}}_{\bm{\gamma}}\right)\right\}^{-1}\left(\widehat{\bm{\gamma}}^{\mathbf{T}} \widehat{\bm{\sigma}}_{\bm{X}}^{-2}\boldsymbol{\mu}_{\bm{\gamma}}\right),
\end{aligned}
\end{equation}
where $\boldsymbol{\mu}_{\bm{\gamma}}=\left(\mu_{\gamma_1},\dots, \mu_{\gamma_J}\right)^{\mathrm{T}}$, $\boldsymbol{\mu}_{\bm{\alpha}}=\left(\mu_{\alpha_1},\dots, \mu_{\alpha_J}\right)^{\mathrm{T}}$, $\boldsymbol{\mathbf{S}
}_{\bm{\gamma}}=\operatorname{diag}\left(\left[\sigma_{\gamma_1}^2,\dots, \sigma_{\gamma_J}^2\right]\right)$ and $\boldsymbol{\mathbf{S}}_{\bm{\alpha}}=\operatorname{diag}\left(\left[\sigma_{\alpha_1}^2,\dots, \sigma_{\alpha_J}^2\right]\right)$. Finally, we use the updated model parameters $\boldsymbol{\theta}$ to construct the evidence lower bound to check the convergence. Since we adopt PX-EM algorithm, the reduction step should be used to process the obtained parameters. More technical details can be found in the Supplementary Materials. 
\\
\indent  After obtaining an estimate of the causal effect, we further calculate the standard error according to the property of likelihood ratio test (LRT) statistics which asymptotically follows the $\chi_{1}^{2}$ under the null hypothesis \citep{van2000asymptotic}. We first formulatey the statistical tests to examine the association between the risk factor and the outcome. 
\begin{equation}
\mathcal{H}_{0}: \beta_{0}=0\qquad \mathcal{H}_{a}: \beta_{0}\neq0 ,
\end{equation}
the likelihood ratio test (LRT) statistics for the causal effect is given by:
\begin{equation}\label{like}
\boldsymbol{ \Lambda}= 2\left(\log \operatorname{Pr}\left(\widehat{\boldsymbol{\gamma}}, \widehat{\Gamma} | \widehat{\bm{\sigma}}_{\bm{X}}, \widehat{\bm{\sigma}}_{\bm{Y}}, \widehat{\bm{\Theta}} ; \boldsymbol{h},\hat{\boldsymbol{\theta}}^{ML}\right) -\log \operatorname{Pr}\left(\widehat{\boldsymbol{\gamma}}, \widehat{\Gamma} | \widehat{\bm{\sigma}}_{\bm{X}}, \widehat{\bm{\sigma}}_{\bm{Y}}, \widehat{\bm{\Theta}} ; \boldsymbol{h},\hat{\boldsymbol{\theta}}_{0}^{ML}\right)\right),
\end{equation}
where $\hat{\boldsymbol{\theta}}_{0}^{ML}$ and $\hat{\boldsymbol{\theta}}^{ML}$ are collections of parameter estimates obtained by maximizing the marginal likelihood under the null hypothesis $\mathcal{H}_0$ and under the alternative hypothesis $\mathcal{H}_a$. We utilize PX-VBEM algorithm to maximize the ELBO to get the $\widehat{\boldsymbol{\theta}}$ and $\widehat{\boldsymbol{\theta}}_{0}$ instead of maximizing the marginal likelihood to overcome the computational intractability. Although PX-VBEM produces accurate posterior mean estimates \citep{blei2017variational,dai2017igess,yang2018lpg}, it would underestimate the marginal variance because we  use the estimated posterior distribution from the ELBO to approximate the marginal likelihood in equation (\ref{like}) \citep{wang2005inadequacy}. Thus, we calibrate ELBO by plugging our estimates ($\widehat{\boldsymbol{\theta}}$ and $\widehat{\boldsymbol{\theta}}_{0}$) from PX-VBEM into the equation (\ref{like}) to construct the test statistics \citep{yang2020comm}:
\begin{equation}\label{other}
\boldsymbol{ \Lambda^{'}}= 2\left(\log \operatorname{Pr}\left(\widehat{\boldsymbol{\gamma}}, \widehat{\Gamma} | \widehat{\bm{\sigma}}_{\bm{X}}, \widehat{\bm{\sigma}}_{\bm{Y}}, \widehat{\bm{\Theta}} ; \boldsymbol{h},\hat{\boldsymbol{\theta}}\right) -\log \operatorname{Pr}\left(\widehat{\boldsymbol{\gamma}}, \widehat{\Gamma} | \widehat{\bm{\sigma}}_{\bm{X}}, \widehat{\bm{\sigma}}_{\bm{Y}}, \widehat{\bm{\Theta}} ; \boldsymbol{h},\hat{\boldsymbol{\theta}}_{0}\right)\right).
\end{equation}
Then, we can get the well-calibrated standard error as  $\widehat{se}(\widehat{\beta_0}$)$=\widehat{\beta_0}/\sqrt{\boldsymbol{ \Lambda^{'}}}$.
\section{Simulation Studies}\label{simulations}
\noindent Although our proposed method is based on GWAS summary level data, we still simulate the individual-level data to better mimic real genetic data sets. Specifically, the data sets are generated according to the following models:
\begin{equation}\label{pp}
\bm{X}=\mathbf{G}_{\bm{X}} \bm{\gamma}+\mathbf{U}_{\bm{X}} \boldsymbol{\eta}_{X}+\bm{\varepsilon}_{\bm{X}}, \quad \bm{Y}=\beta_{0} \mathbf{X}+\mathbf{G}_{\bm{Y}} \boldsymbol{\alpha}+\mathbf{U}_{\bm{Y}} \boldsymbol{\eta}_{Y}+\bm{\varepsilon}_{\bm{Y}},
\end{equation}
where $\bm{X}\in \mathbb{R}^{n_X \times 1}$ is the exposure vector,  $\bm{Y}\in \mathbb{R}^{n_Y \times 1}$ is the outcome vector, $\mathbf{G}_{\bm{X}} \in \mathbb{R}^{n_{X} \times J}$ and $\mathbf{G}_{\bm{Y}} \in \mathbb{R}^{n_{Y} \times J}$ are the genotype datasets for the exposure $\bm{X}$ and the outcome $\bm{Y}$, $\mathbf{U}_{\bm{X}} \in \mathbb{R}^{n_{X} \times N_0}$ and $\mathbf{U}_{\bm{Y}} \in \mathbb{R}^{n_{Y} \times N_0}$ are matrices for confounding variables, $n_X$ and $n_Y$ are the corresponding sample sizes of exposure $\bm{X}$ and outcome $\bm{Y}$, $J$ is the number of genotyped SNPs. The error terms  $\bm{\varepsilon}_{\bm{X}}$ and $\bm{\varepsilon}_{\bm{Y}}$ are independent noises generated from $\mathcal{N}\left(\mathbf{0}, {\sigma}_{\bm{\varepsilon}_{\bm{X}}}^{\mathbf{2}} \mathbf{I}_{\bm{n_X}}\right)$ and $\mathcal{N}\left(\mathbf{0}, {\sigma}_{\bm{\varepsilon}_{\bm{Y}}}^{\mathbf{2}} \mathbf{I}_{\bm{n_Y}}\right)$, where the values of ${\sigma}_{\bm{\varepsilon}_{\bm{X}}}^{\mathbf{2}}$  and ${\sigma}_{\bm{\varepsilon}_{\bm{Y}}}^{\mathbf{2}}$  are around 0.8 and 0.4 on average, respectively. In model (\ref{pp}), $\beta_0$ is the true causal effect and $\bm{\alpha}$ represents the direct effect of the SNPs on the outcome not mediated by the exposure variable, where $\alpha_{j} \stackrel{i . i . d}{\sim}\mathcal{N}\left(0, \sigma_{0}^{2}\right)$,  $j=1,2,\dots,500$. To simulate the idiosyncratic pleiotropy, we randomly select 5$\%$ of IVs so that their direct effect $\alpha_j$s have mean 0 and standard deviation $40\sigma_0$, where $\sigma_0^2 = 0.008$.\\
\indent An external reference panel $\mathbf{G}_r\in  \mathbb{R}^{n_{r} \times J}$ is chosen for estimating the LD matrix among SNPs, where $n_r = 5000$ is the sample size of the chosen reference panel. We used the R package {\it MR.LDP} to generate the genotype matrices $\mathbf{G}_{\bm{X}}$, $\mathbf{G}_{\bm{Y}}$ and $\mathbf{G}_{\bm{r}}$ by mimicking the LD structure in the CAD-CAD data set as in Section \ref{real}. We fix $n_X=n_Y=20000$. The number of blocks is set to be 10 and the number of SNPs within each block is 50. Thus, the  total number of SNPs is $J=500$.   The confounders are generated as follows:
 \begin{align}
     \mathbf{U}_{\bm{X}} &= \mathbf{G}_{\bm{X}}\bm{\phi}_{\bm{X}} + \bm{\xi}_{\mathbf{U}_{\bm{X}}},\\
     \mathbf{U}_{\bm{Y}} &= \mathbf{G}_{\bm{Y}}\bm{\phi}_{\bm{Y}} + \bm{\xi}_{\mathbf{U}_{\bm{Y}}}.
 \end{align} 
 Each row of $\bm{\phi}_{\bm{X}}$ and $\bm{\phi}_{\bm{Y}}$ is sampled from $\mathcal{N}\left(\mathbf{0}, {\sigma}_{\bm{\phi}_{\bm{X}}}^{\mathbf{2}} \mathbf{I}_{\bm{n_X}}\right)$ and $\mathcal{N}\left(\mathbf{0}, {\sigma}_{\bm{\phi}_{\bm{Y}}}^{\mathbf{2}} \mathbf{I}_{\bm{n_Y}}\right)$, where ${\sigma}_{\bm{\phi}_{\bm{X}}} = {\sigma}_{\bm{\phi}_{\bm{Y}}} = 0.01$, respectively. We sample each column of $\bm{\xi}_{\mathbf{U}_{\bm{X}}}$ and $\bm{\xi}_{\mathbf{U}_{\bm{Y}}}$ from a standard normal distribution,
 while each row of the corresponding coefficients $\boldsymbol{\eta}_{\bm{X}} \in \mathbb{R}^{N_0 \times 1}$ and $\boldsymbol{\eta}_{\bm{Y}} \in \mathbb{R}^{N_0 \times 1}$ of the confounders is sampled from a multivariate normal distribution $\mathcal{N}\left(\mathbf{0}, \bm{S}_{\bm{\eta}}\right)$ where the diagonal elements of $\bm{S}_{\bm{\eta}}\in \mathbb{R}^{2 \times 2}$ are 1 and the off-diagonal  elements are 0.85.
\begin{figure}[htbp]
\centering
\captionsetup[subfloat]{captionskip=5pt}
\subfloat[]{
\label{fig:subfig_a}
\begin{minipage}{0.5\linewidth}
\centering
\includegraphics[width = 8cm, height = 7cm]{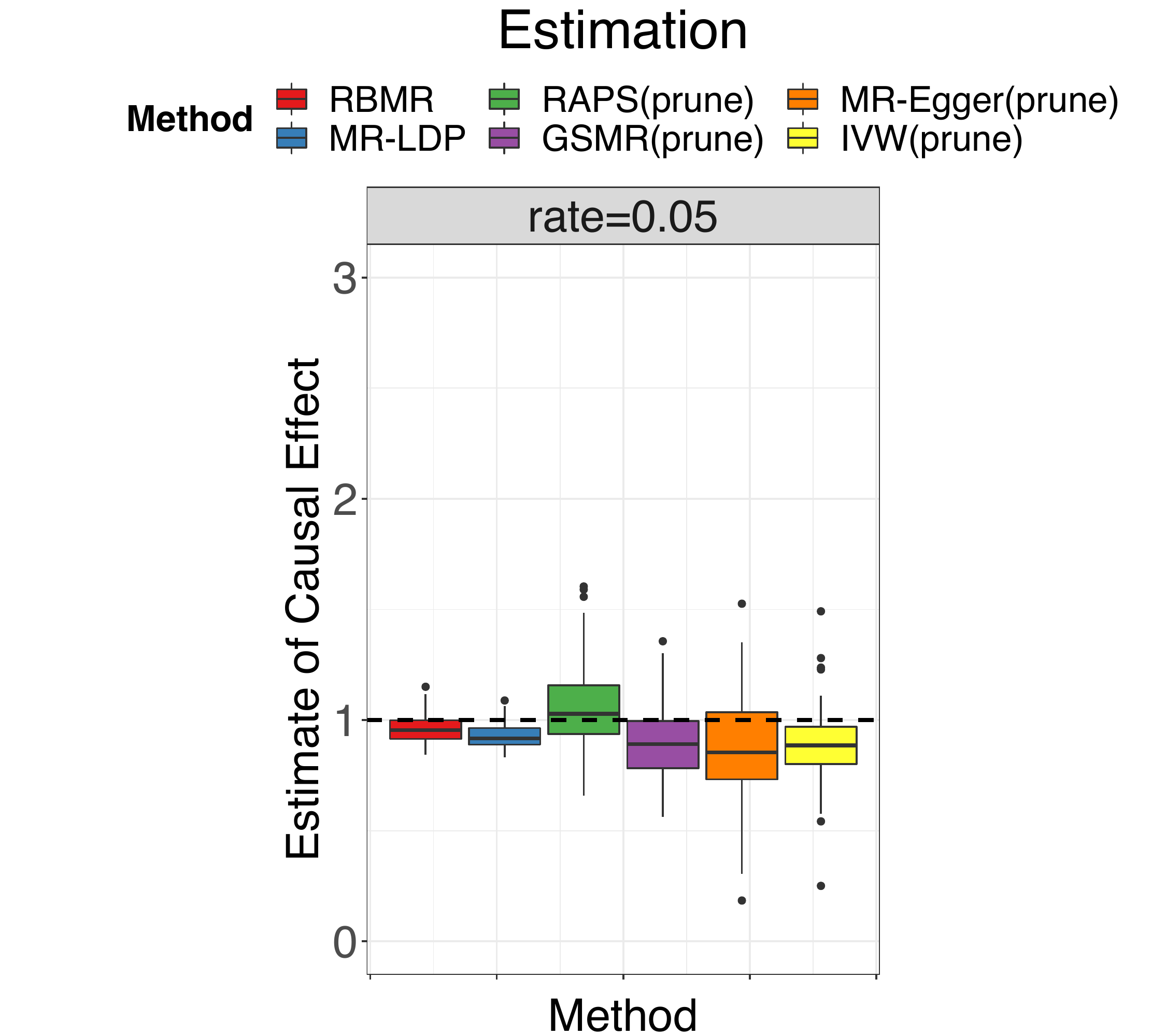}
\end{minipage}
}
\subfloat[]{
\label{fig:subfig_b}
\begin{minipage}{0.5\linewidth}
\centering
\includegraphics[width = 8cm, height = 7cm]{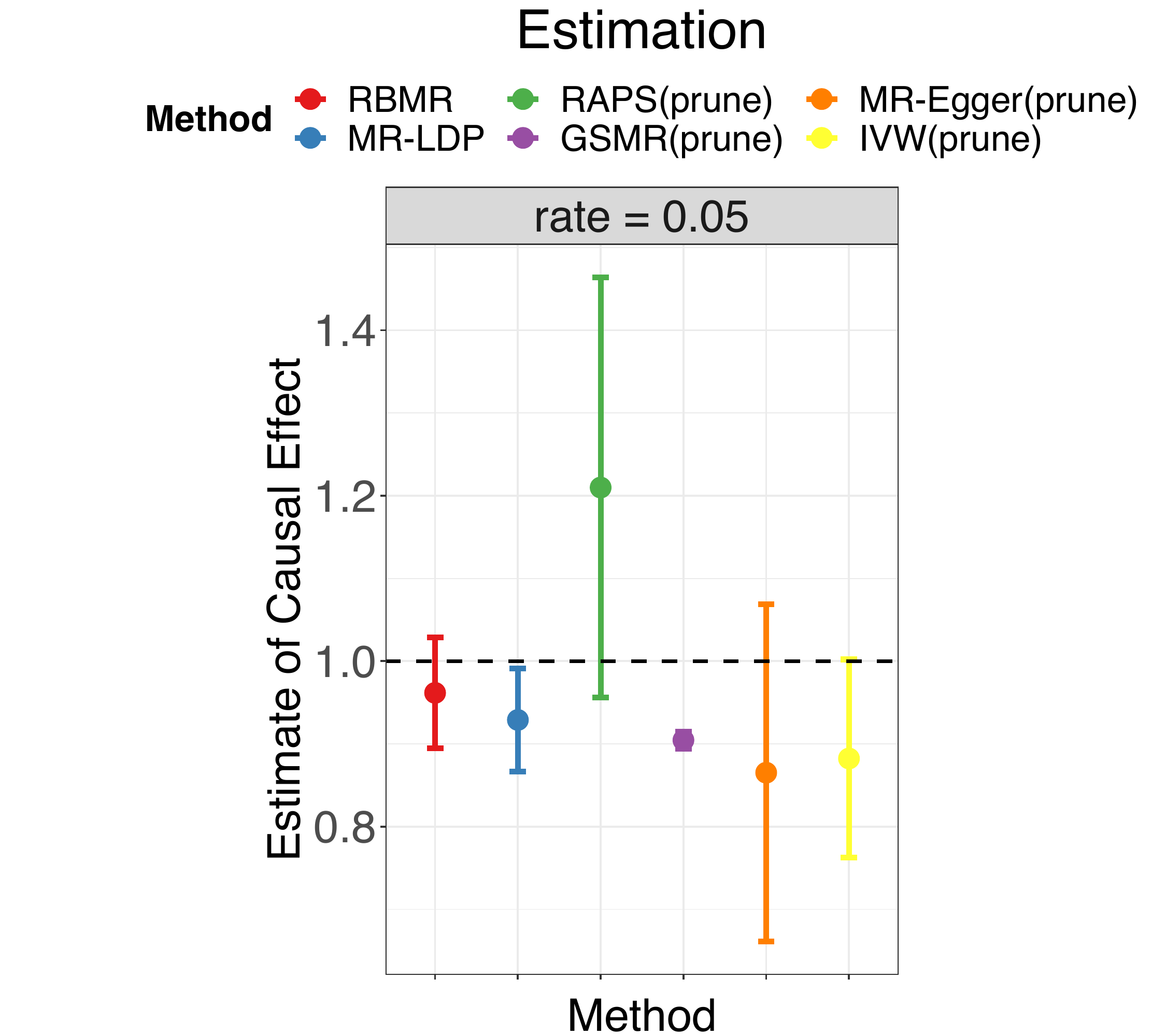}
\end{minipage}
}

\vspace{10pt}
 \subfloat[]{
\label{fig:subfig_c}
\begin{minipage}{\linewidth}
\centering
\includegraphics[width = 12cm, height = 7cm]{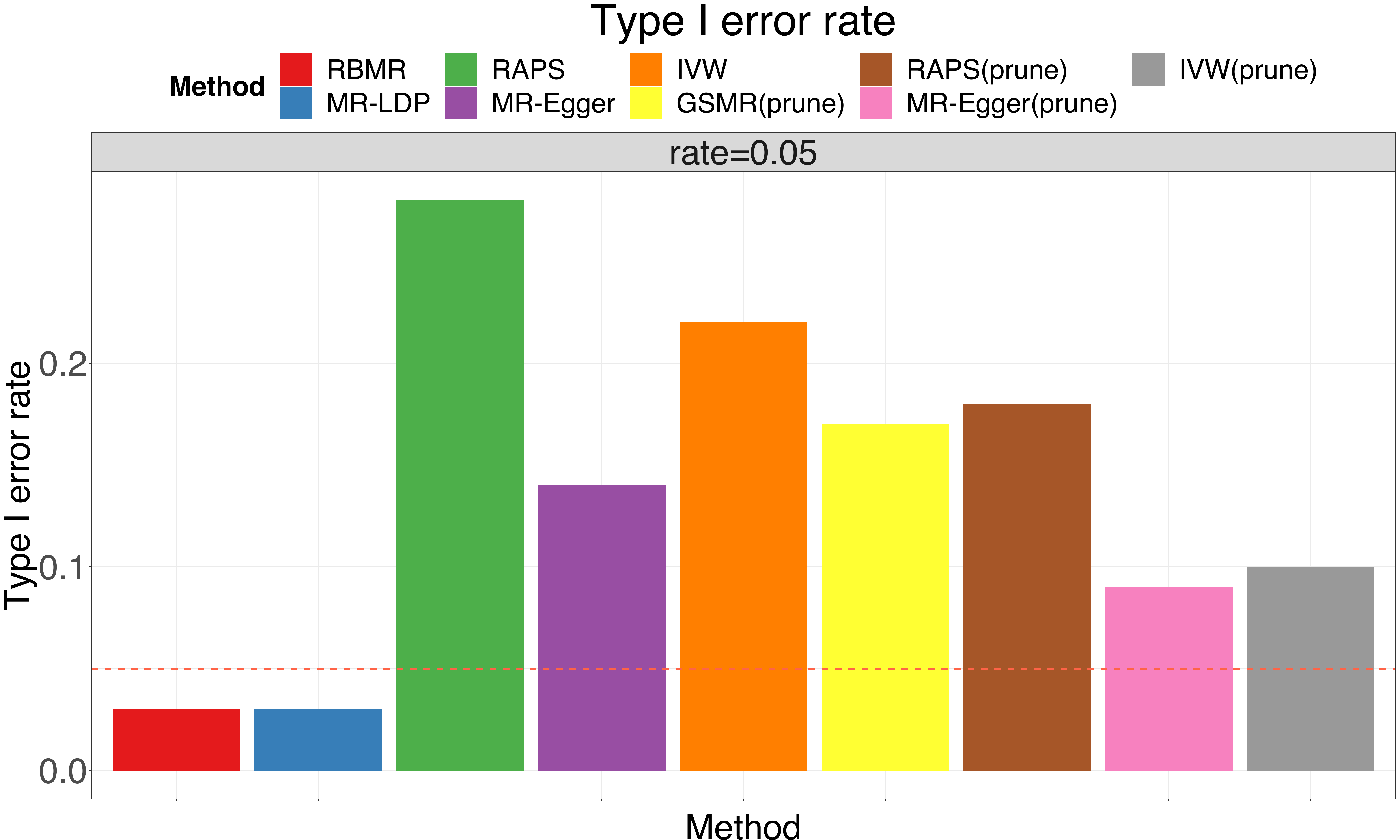}
\end{minipage}
}
\captionsetup{font+=small}
\caption{Comparisons of MR methods affected by the LD and pleiotropy. Figure (a) is a boxplot, Figure (b) contains point estimates and 95\% confidence intervals,  Figure (c) contains all the type I error rates of the methods.}
\label{fig:Figure1}
\end{figure}
\begin{table}[htbp]
\centering     
\captionsetup{font+=small}
\caption{Comparisons of the point estimates in the terms of bias\%, RMSE\% and the coverage probabilities.}
\resizebox{0.6\textwidth}{!}{\begin{tabular}{ccccc}\hline
     Method           & $\widehat{\beta}$ & Bias\%  & RMSE\%  & Cover\% \\
     \hline
RBMR             & 0.962     & -3.837  & 7.583   & 94.000  \\
MR-LDP           & 0.929     & -7.107  & 8.834   & 86.000  \\
GSMR (prune)     & 0.905     & -9.549  & 18.728  & 5.000   \\
RAPS (prune)     & 1.210     & 20.981  & 155.223 & 87.000  \\
MR-Egger (prune) & 0.865     & -13.484 & 27.689  & 86.000  \\
IVW (prune)      & 0.882     & -11.768 & 19.996  & 73.000 \\
\hline
      \end{tabular}}
      \end{table}

\noindent The signal magnitude for $\bm{\alpha}$ 
is controlled by the heritability $h_{\bm{\alpha}}$ due to systematic pleiotropy,  $h_{\bm{\alpha}}^{{2}}=\frac{\operatorname{var}\left(\mathbf{G}_{\bm{Y}} \boldsymbol{\alpha}\right)}{\operatorname{var}(\bm{Y})} = 0.05$. The signal magnitude for $\bm{\gamma}$ is chosen such that the heritability $h_{\bm{\gamma}}^{2}=\frac{\operatorname{var}\left(\beta_{0} \mathbf{G}_{\bm{X}} \bm{\gamma}\right)}{\operatorname{var}(\bm{Y})}=0.1$. Then we control the heritability for $\bm{X}$ at 0.1. The true causal effect $\beta_0$ is set to be 1.\\
\indent We first run single-variant genetic association analysis for the exposure and the outcome respectively, and then we obtain the summary-level statistics $\{\widehat{\gamma}_j,\widehat{\Gamma}_j\}_{j=1,2,\dots,500}$ with their corresponding standard errors $\{\widehat{\sigma}_{X_j},\widehat{\sigma}_{Y_j}\}_{j=1,2,\dots,500}$.  Then we use the summary-level data to conduct MR analyses using the proposed RBMR, MR-LDP, MR-Egger, RAPS, GSMR and IVW methods. As the prerequisite for MR-Egger, RAPS and IVW methods is that the instrumental variables are independent of each other, we perform LD pruning by controlling the LD $r^2$ at the threshold 0.05 \citep{zhu2018causal}. We repeat the simulations for 500 times.\\
\indent  We evaluate the  type-I error rates under the null that $\beta_0 = 0$ and evaluate the estimation accuracy of point estimates under the alternative that $\beta_0 = 1$. Figure 1 shows the type-I error rates and point estimates for all the methods. As shown in Figure 1(c), the proposed RBMR and MR-LDP methods control the type-I errors at the nominal level 0.05. Although after LD pruning, genetic variants are independent, however, the competing methods, GSMR, RAPS, MR-Egger and IVW still fail to control the type-I error because of the presence of idiosyncratic pleiotropy. We found that our method RBMR and MR-LDP are more stable than the other four methods as shown in Figure 1(a). But we found that our method RBMR is more accurate than MR-LDP in terms of  relative bias, root mean square error (RMSE\%) and coverage probabilities as shown in Figure 1(b) and Table 1.   We conducted more simulation studies and obtain essentially the same conclusion. Detailed results are provided  in the Supplementary Materials.   

\section{Real Data Analysis}\label{real}
\noindent In this section, we analyzed four real data sets to demonstrate the performance of our proposed method. The 1000 Genome Project Phase 1 (1KGP) is used as the reference panel  to compute the LD matrix \citep{10002012integrated}. We first analyze two benchmark data sets commonly used for method comparison purpose, then we will estimate the causal effect of coronary artery disease (CAD) on the risk of critically ill COVID-19 outcome defined as those who end up on respiratory support or die from COVID-19. We also estimate the causal effect of low-density lipoprotein (LDL) cholesterol on the risk of Alzheimer's disease. \\
\indent The first  benchmark data analysis is based on the summary-level data sets from two non-overlapping GWAS studies for the  coronary artery disease (CAD), usually referred to as the CAD-CAD data.  The true causal effect should be exactly one.  
The selection data set is from the Myocardial Infarction Genetics in the UK Biobank , the exposure data is from the Coronary Artery Disease (C4D) Genetics Consortium \citep{coronary2011genome}, and  the outcome data is from the transatlantic Coronary Artery Disease Genome Wide Replication and Meta-analysis (CARDIoGRAM) \citep{schunkert2011large}. We first filter the genetic variants using the selection data under different association $p$-value thresholds ($p$-value $\le 1\times10^{-4}, 5\times 10^{-4},1\times 10^{-3})$.  Then we applied our proposed RBMR method and the MR-LDP to all the selected and possibly correlated SNPs by accounting for the LD structure explicitly. We  applied the GSMR, IVW, MR-Egger and MR-RAPS methods using the independent  SNPs after LD pruning at the LD threshold 0.05.  We obtain causal effect point estimates and the corresponding 95\% confidence intervals (CI) as shown in Figure 2(a). We found that our proposed RBMR method outperforms other methods because it has the smallest bias and shortest confidence intervals for a range of $p$-value thresholds. Our proposed method RBMR used all selected SNPs (without LD pruning) in the selection data set and thus we might obtain more accurate causal effect estimate. However, other methods might be biased due to the pruning process, because the pruning process might filter out the `good' IVs and keep the `bad' IVs.\\
\indent To further investigate the performance of our proposed RBMR method, we consider the case that both the exposure and outcome are body mass index (BMI). We select SNPs based on previous research \citep{locke2015genetic}. The exposure  is the BMI for physically active men  and the outcome  is  the BMI for physically active women, both are of European ancestry (\url{https://portals.broadinstitute.org/collaboration/giant/index.php/GIANT_consortium_data_files#2018_GIANT_and_UK_BioBank_Meta_Analysis_for_Public_Release}). The point estimates and the corresponding 95\% confidence intervals  are shown in Figure 2(b). We found that our proposed RBMR method has smaller bias than other competing methods. More numerical results are provided in the  Supplementary Materials.\\
\begin{figure}[]
\centering
\captionsetup[subfloat]{captionskip=5pt}
\subfloat[CAD-CAD]{
\label{fig:subfig_h}
\hspace{-1.2cm}
\begin{minipage}[b]{.5\linewidth}
\centering
\includegraphics[width=8.5cm,height=6.5cm]{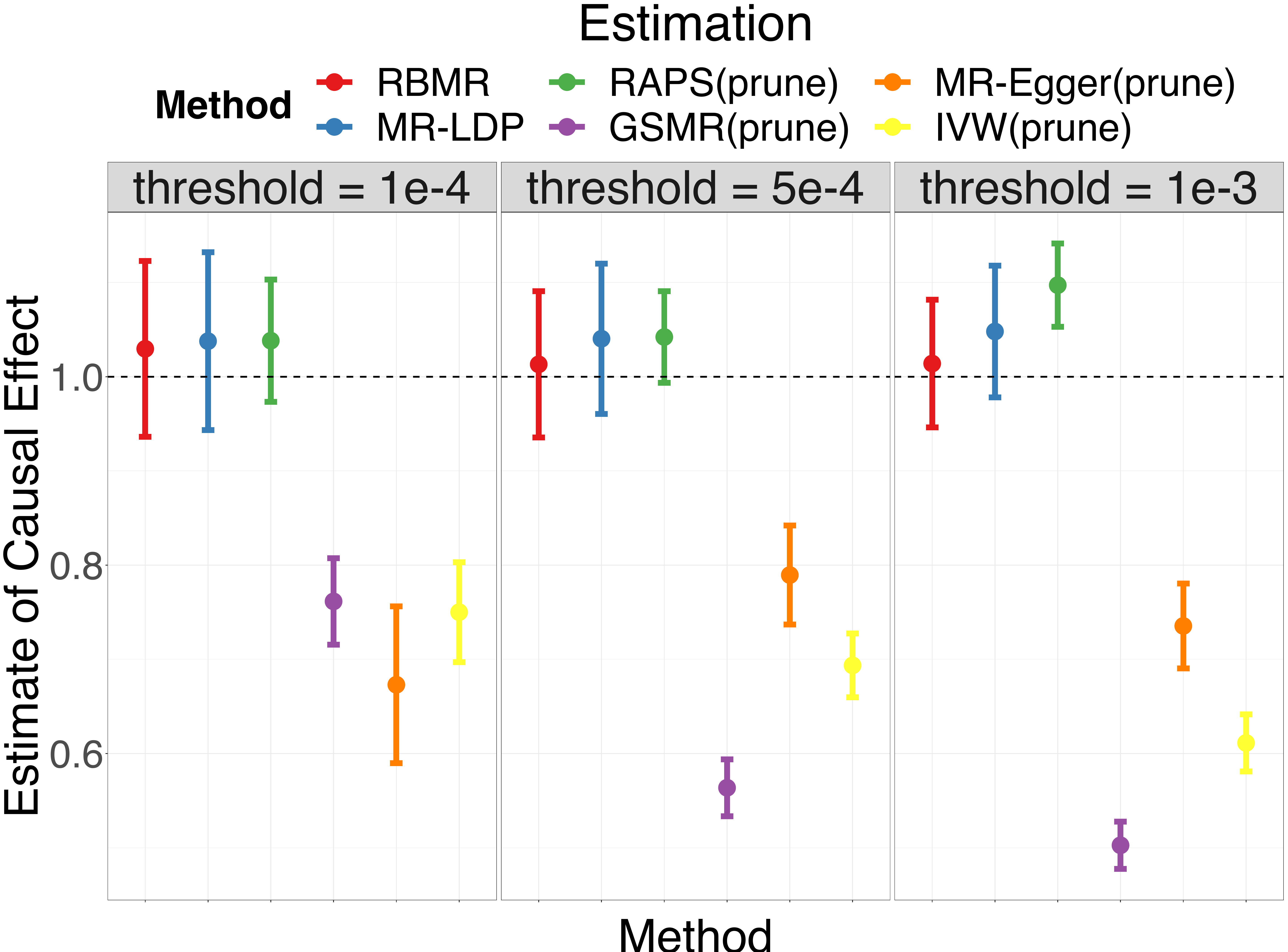}
\end{minipage}
}
\subfloat[BMI-BMI]{
\label{fig:subfig_k}
\hspace{1cm}
\begin{minipage}[b]{.5\linewidth}
\centering
\includegraphics[width=8.5cm,height=6.5cm]{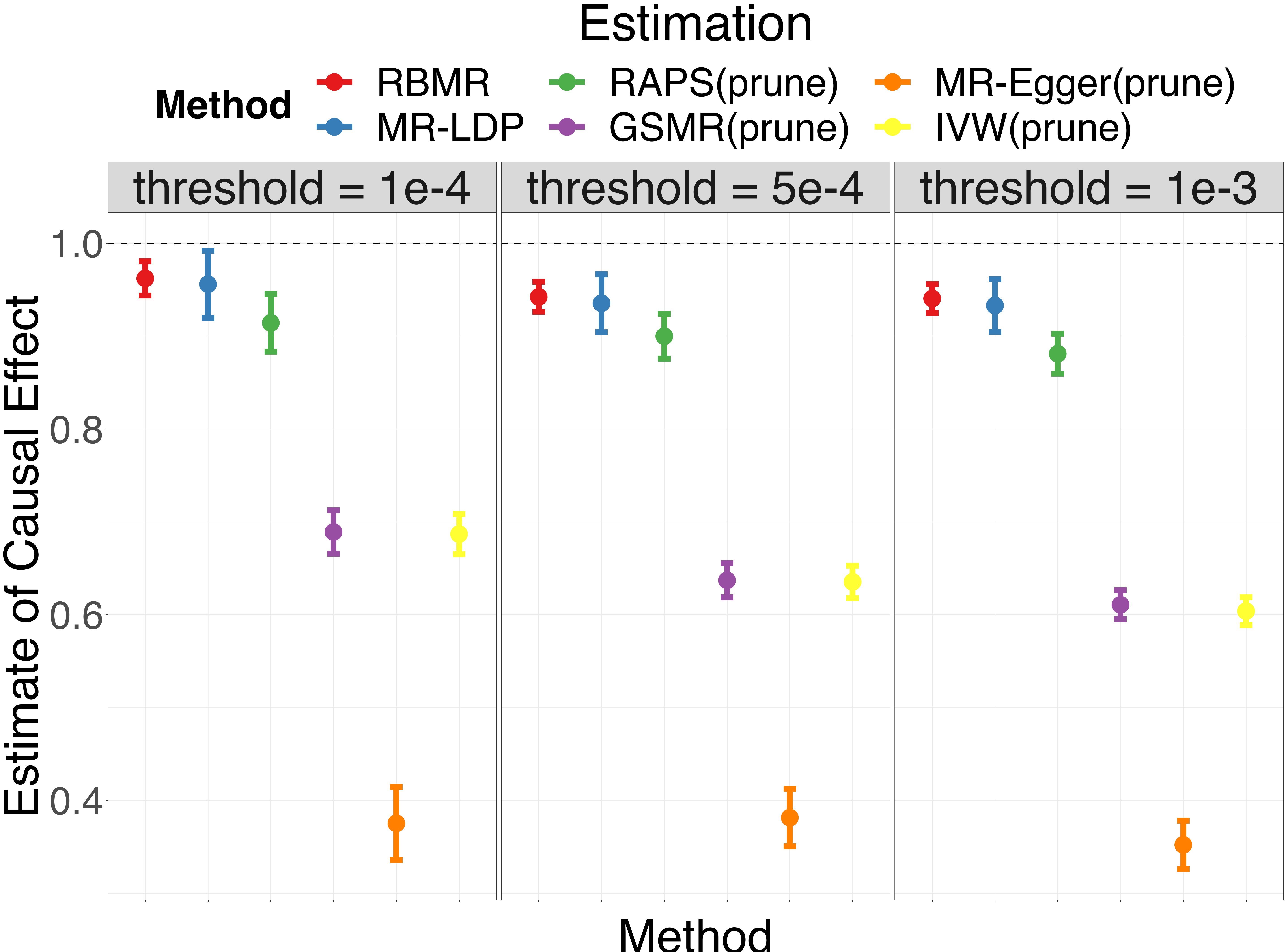}
\end{minipage}
}
\caption{The results of CAD-CAD and BMI-BMI using 1KGP as the reference panel with shrinkage parameter $\lambda=0.15$. The SNPs are selected at the three thresholds ($p$-value $\le 1\times10^{-4}, 5\times 10^{-4},1\times 10^{-3})$.}
\label{fig:Figure7}
\end{figure}
\indent We apply our proposed RBMR method together with other competing methods to estimate the causal effect of CAD on the risk of critically ill coronavirus disease 2019 (COVID-19) defined as those who end up on respiratory support or die from COVID-19. Specifically, the selection data set is the Myocardial Infraction Genetics in the UK Biobank and the exposure data set is from \citet{coronary2011genome}. The outcome is obtained from Freeze 5 (January 2021) of the COVID-19 Host Genetics Initiative (COVID-19 HGI) Genome-Wide Association Study \citep{covid2020covid} (\url{https://www.covid19hg.org/results/}). The data combines the  genetic data of 49562 patients and two million controls from 46 studies across 19 countries \citep{covid2021mapping}.   We mainly consider the GWAS data on the 6179 cases with critical illness due to COVID-19 and 1483780 controls from the general populations in our analysis. We  use the selection data with  $p$-value $\le 1\times 10^{-4}$ threshold to select  genetic variants as IVs.   As shown in Figure 3(a), we found a significant effect of CAD on the risk of critically ill COVID-19 using our RBMR method ($\widehat{\beta}=0.261$, $p$-value =
0.008, 95$\%$ CI = (0.067, 0.454)), MR-LDP ($\widehat{\beta}=0.258$, $p$-value = 
0.009, 95$\%$ CI = (0.065, 0.452)), GSMR ($\widehat{\beta}=0.201$, $p$-value = 0.045, 95$\%$ CI = (0.004, 0.398)), MR-Egger ($\widehat{\beta}=0.313$, $p$-value = 0.036, 95\% CI = (0.020,  0.605)) and IVW ($\widehat{\beta}=0.201$, $p$-value = 0.045, 95\% CI = (0.005,  0.397)) . However, the result of GSMR ($\widehat{\beta}=0.268$, $p$-value = 0.073, 95$\%$ CI = (-0.025, 0.561)) is  not significant ($p$-value $>$ 0.05).
Our RBMR is more accurate as its confidence interval is slightly shorter and its $p$-value is more significant.   \\
\begin{figure}[]
\subfloat[CAD-COVID-19]{
\label{fig:subfig_m}
\hspace{-2.2cm}\begin{minipage}[b]{0.45\linewidth}
\centering
\includegraphics[width=10cm,height=7cm]{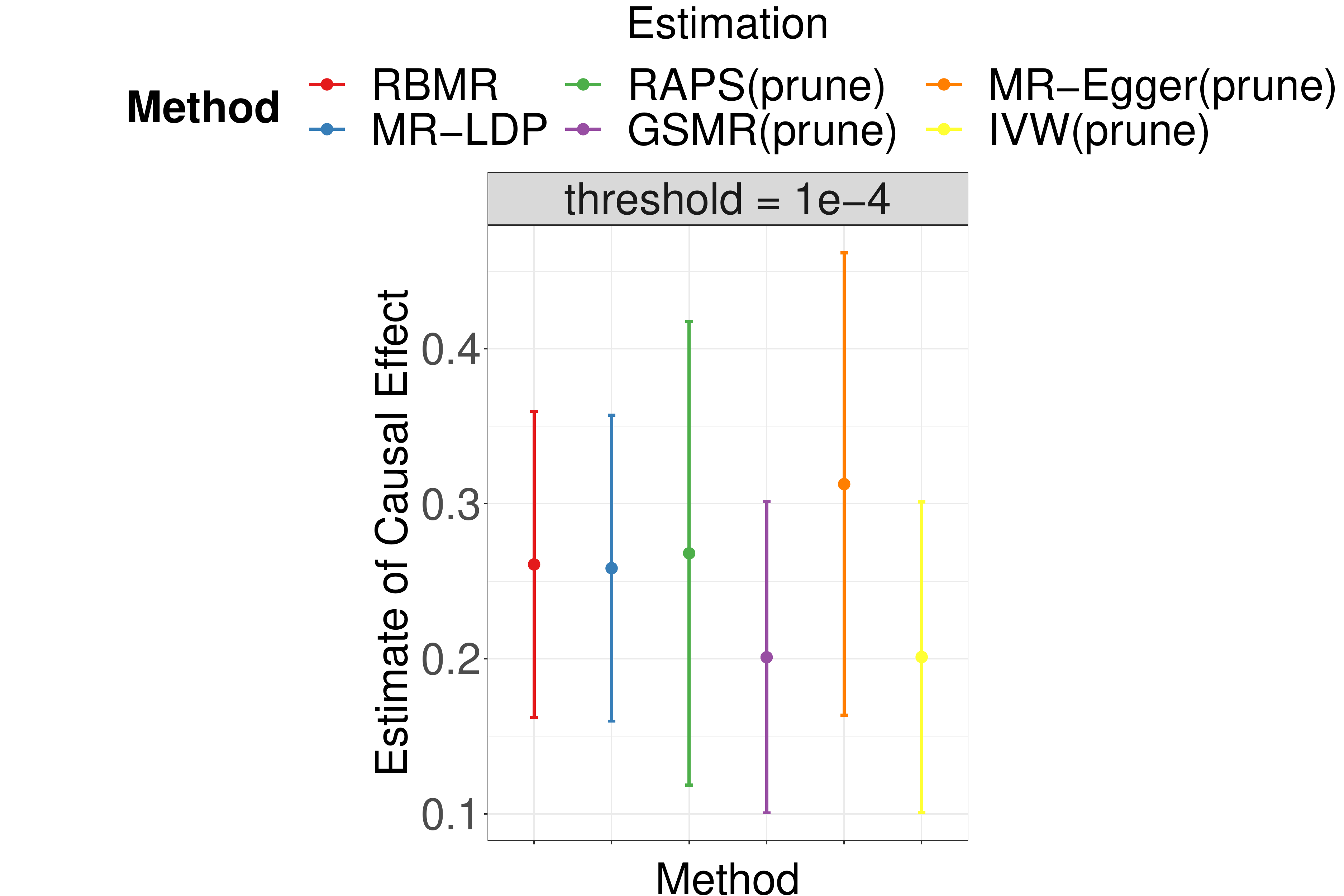}
\end{minipage}
}
\subfloat[Scatter plot of $\widehat{\Gamma_j}$ versus $\widehat{\gamma_j}$ for CAD-COVID-19 data]{
\label{fig:subfig_n}
\hspace{2.2cm}
\begin{minipage}[b]{0.65\linewidth}
\centering
\includegraphics[width=10cm,height=7cm]{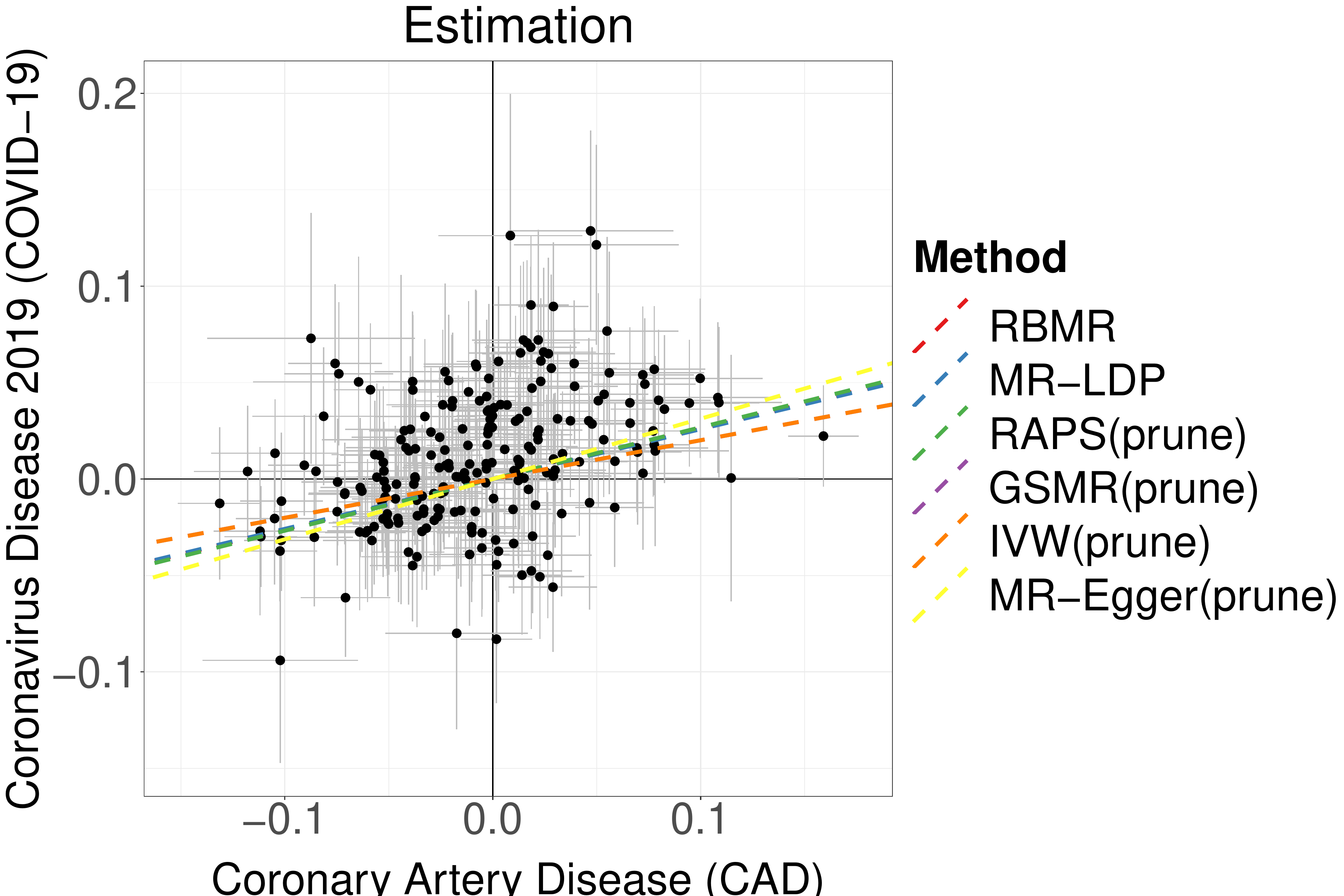}
\end{minipage}
}
\caption{The results of CAD-COVID-19 using 1KGP as the reference panel with shrinkage parameter $\lambda=0.1$. The SNPs are selected  at the threshold $($$p$-value $\le 1\times10^{-4}$$)$. Each point of the scatter plot in Figure \protect\subref{fig:subfig_n} is augmented by the standard errors of $\widehat{\Gamma_j}$ and  $\widehat{\gamma_j}$ on the vertical and horizontal sides respectively. Dashed lines are the slopes fitted by the six methods.}
\label{fig:Figure8}
\end{figure}
 \indent We further  apply our proposed RBMR and other competing methods to estimate  the causal effect of  LDL cholesterol on the risk of Alzheimer's disease. The selection data set is from \citet{teslovich2010biological} with 95454 individuals, and the exposure data set  is  from \citet{willer2013discovery} with 188577 individuals. The outcome data set is obtained from the stage 1 meta-analysis of four GWAS samples (n = 54612) of the International Genomics of Alzheimer's Project \citep{lambert2013meta}. We select the SNPs at the $p$-value threshold $5\times 10^{-8}$. The results are summarized in Figure 4. We find that the causal effect of RBMR is $\widehat{\beta} = 0.122$ ($p$-value = $1.156 \times 10^{-3}$, 95$\%$ CI = (0.048, 0.196)), the estimate of MR-LDP is $\widehat{\beta} = 0.232$ ($p$-value = $1.242\times 10^{-3}$, 95$\%$ CI = (0.091, 0.374))  and the estimate of RAPS is $\widehat{\beta} = 0.152$ ($p$-value = $5.506\times 10^{-7}$, 95$\%$ CI = (0.093, 0.212)). The estimates of  IVW ($\widehat{\beta} = 0.858$, $p$-value = $6.648 \times 10^{-7}$, 95$\%$ CI = (0.520, 1.196)) and the MR-Egger ($\widehat{\beta} = 1.472$, $p$-value = $1.574 \times 10^{-6}$, 95$\%$ CI = (0.871 2.072)) are much larger than the estimates of RBMR, MR-LDP and RAPS. And the estimate of GSMR ($\widehat{\beta} = 0.033$, $p$-value = 0.393, 95$\%$ CI = (-0.043 0.109)) is much smaller than the estimates of RBMR, MR-LDP and RAPS. Since there exists obvious idiosyncratic pleiotropy in this data set, hence the estimates of IVW, MR-Egger and GSMR are likely to be biased. Both RAPS and MR-LDP use the normal distribution to model the direct effects which might be violated in the presence of the idiosyncratic pleiotropy as in this data set, therefore the estimates of RAPS and MR-LDP might have upward bias. 
 \begin{figure}[htb]
\centering
\includegraphics[width=12cm,height=8cm]{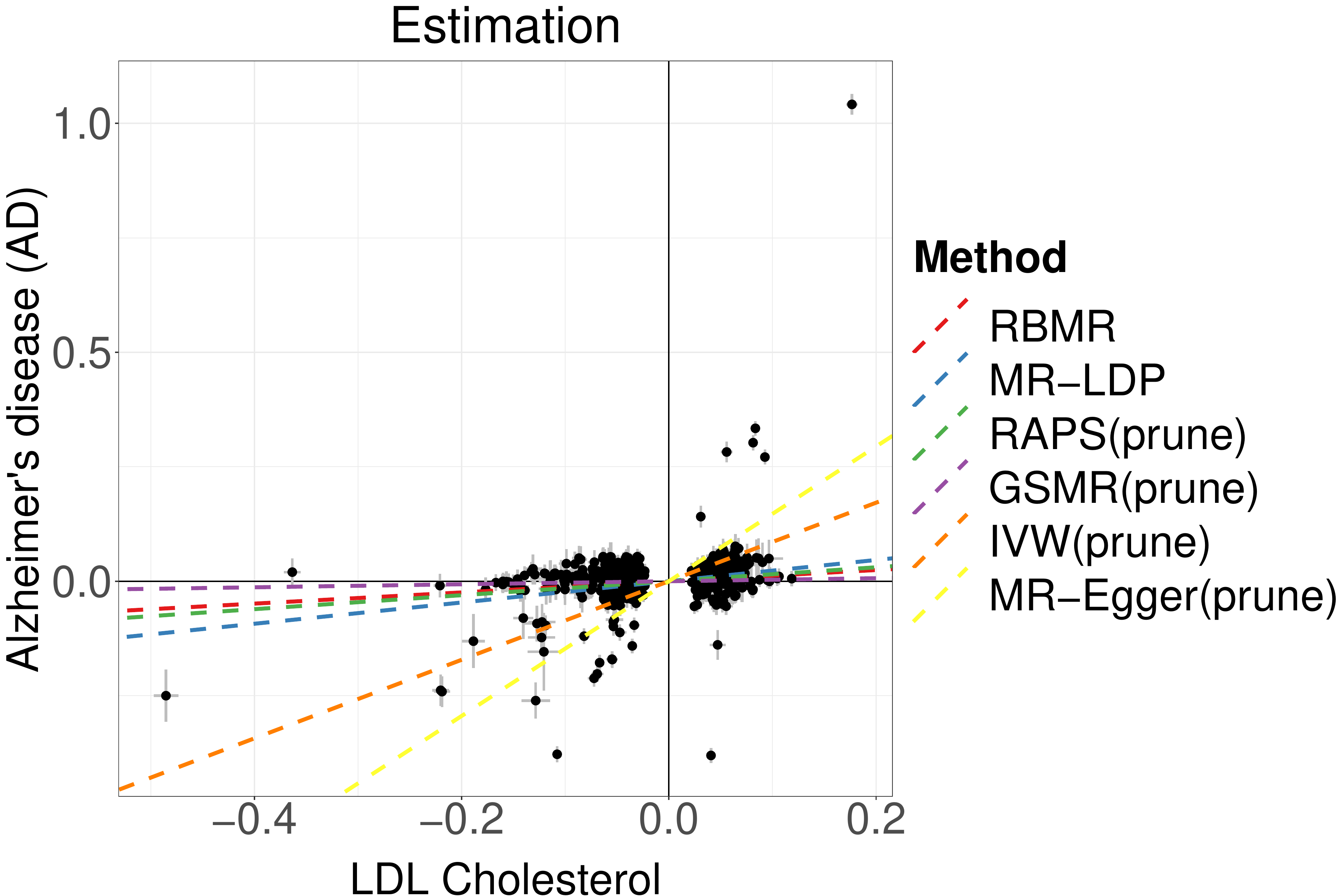}

\caption{The results of LDL cholesterol on Alzheimer’s disease using 1KGP as the reference panel with shrinkage parameter $\lambda=0.15$. The SNPs are selected at the threshold (p-value $\le 5\times10^{-8}$). Each point of the scatter plot is augmented by the standard errors of $\widehat{\Gamma_j}$ and  $\widehat{\gamma_j}$ on the vertical and horizontal sides respectively. Dashed lines are the slopes fitted by the six methods.}
\label{fig:Figure9}
\end{figure}
\section{Discussion}\label{Discussion}
\noindent In this paper, we propose a novel two-sample robust MR method RBMR by accounting for the LD structure, systematic pleiotropy and idiosyncratic pleiotropy simultaneously in a unified framework.  Specifically, we propose to use the more robust multivariate generalized $t$-distribution rather the less robust Gaussian distribution to model the direct effects of the IV on the outcome not mediated by the exposure. Moreover,  the multivariate generalized $t$-distribution can be reformulated as Gaussian scaled mixtures to facilitate the estimation of the model parameters using the parameter expanded variational Bayesian expectation-maximum algorithm (PX-VBEM).  Through extensive simulations and analysis of two real benchmark data sets, we found that our method outperforms the other competing methods. We find that CAD might increase the risk of critically ill COVID-19, and higher level of LDL cholesterol might increase the risk of Alzheimer's disease.\\
\indent We make the following two major contributions. First, our method can account for the LD structure explicitly and thus can include more possibly correlated SNPs to reduce bias and increase estimation efficiency. Second, our RBMR method is more robust to the presence of idiosyncratic pleiotropy. This enhanced robustness can be very helpful in practice as shown by our simulation studies and real data analysis.  One limitation of our proposed method is that  it cannot handle correlated pleiotropy where the direct effect of the IV on the outcome might be correlated with the IV strength. We leave it as our future work.

\section*{Acknowledgements}
\noindent Dr. Zhonghua Liu’s research is supported by the Start-up research fund (000250348) of the University of Hong Kong and Guangdong Natural Science Fund (2021A1515010268). The authors also thank the editors and reviewers for their constructive comments.

\bibliographystyle{apalike} \bibliography{RBMRrevision}

\begin{thebibliography}{}

\bibitem[Ala-Luhtala and Pich{\'e}, 2016]{ala2016gaussian}
Ala-Luhtala, J. and Pich{\'e}, R. (2016).
\newblock Gaussian scale mixture models for robust linear multivariate
  regression with missing data.
\newblock {\em Communications in Statistics-Simulation and Computation}, 45(3).

\bibitem[Arellano-Valle and Bolfarine, 1995]{arellano1995some}
Arellano-Valle, R.~B. and Bolfarine, H. (1995).
\newblock On some characterizations of the t-distribution.
\newblock {\em Statistics \& Probability Letters}, 25(1):79--85.

\bibitem[Beal et~al., 2003]{beal2003variational}
Beal, M.~J. et~al. (2003).
\newblock {\em Variational algorithms for approximate Bayesian inference}.
\newblock University of London London.

\bibitem[Berisa and Pickrell, 2016]{berisa2016approximately}
Berisa, T. and Pickrell, J.~K. (2016).
\newblock Approximately independent linkage disequilibrium blocks in human
  populations.
\newblock {\em Bioinformatics}, 32(2):283.

\bibitem[Blei et~al., 2017]{blei2017variational}
Blei, D.~M., Kucukelbir, A., and McAuliffe, J.~D. (2017).
\newblock Variational inference: A review for statisticians.
\newblock {\em Journal of the American Statistical Association},
  112(528):859--877.

\bibitem[Bound et~al., 1995]{bound1995problems}
Bound, J., Jaeger, D.~A., and Baker, R.~M. (1995).
\newblock Problems with instrumental variables estimation when the correlation
  between the instruments and the endogenous explanatory variable is weak.
\newblock {\em Journal of the American Statistical Association},
  90(430):443--450.

\bibitem[Bowden et~al., 2015]{bowden2015mendelian}
Bowden, J., Davey~Smith, G., and Burgess, S. (2015).
\newblock Mendelian randomization with invalid instruments: effect estimation
  and bias detection through {Egger} regression.
\newblock {\em International Journal of Epidemiology}, 44(2):512--525.

\bibitem[Burgess et~al., 2013]{burgess2013mendelian}
Burgess, S., Butterworth, A., and Thompson, S.~G. (2013).
\newblock Mendelian randomization analysis with multiple genetic variants using
  summarized data.
\newblock {\em Genetic Epidemiology}, 37(7):658--665.

\bibitem[Cheng et~al., 2020]{cheng2020mr}
Cheng, Q., Yang, Y., Shi, X., Yeung, K.-F., Yang, C., Peng, H., and Liu, J.
  (2020).
\newblock {MR-LDP}: a two-sample mendelian randomization for gwas summary
  statistics accounting for linkage disequilibrium and horizontal pleiotropy.
\newblock {\em NAR Genomics and Bioinformatics}, 2(2):lqaa028.

\bibitem[Consortium et~al., 2012]{10002012integrated}
Consortium, . G.~P. et~al. (2012).
\newblock An integrated map of genetic variation from 1,092 human genomes.
\newblock {\em Nature}, 491(7422):56.

\bibitem[Consortium et~al., 2011]{coronary2011genome}
Consortium, C. A. D. C.~G. et~al. (2011).
\newblock A genome-wide association study in europeans and south asians
  identifies five new loci for coronary artery disease.
\newblock {\em Nature Genetics}, 43(4):339.

\bibitem[Dai et~al., 2017]{dai2017igess}
Dai, M., Ming, J., Cai, M., Liu, J., Yang, C., Wan, X., and Xu, Z. (2017).
\newblock {IGESS}: a statistical approach to integrating individual-level
  genotype data and summary statistics in genome-wide association studies.
\newblock {\em Bioinformatics}, 33(18):2882--2889.

\bibitem[Dempster et~al., 1977]{dempster1977maximum}
Dempster, A.~P., Laird, N.~M., and Rubin, D.~B. (1977).
\newblock Maximum likelihood from incomplete data via the em algorithm.
\newblock {\em Journal of the Royal Statistical Society: Series B
  (Methodological)}, 39(1):1--22.

\bibitem[Ebrahim and Smith, 2008]{ebrahim2008mendelian}
Ebrahim, S. and Smith, G.~D. (2008).
\newblock Mendelian randomization: can genetic epidemiology help redress the
  failures of observational epidemiology?
\newblock {\em Human Genetics}, 123(1):15--33.

\bibitem[Ehret et~al., 2011]{ehret2011genetic}
Ehret, G.~B., Munroe, P.~B., Rice, K.~M., Bochud, M., Johnson, A.~D., Chasman,
  D.~I., Smith, A.~V., Tobin, M.~D., Verwoert, G.~C., Hwang, S.-J., et~al.
  (2011).
\newblock Genetic variants in novel pathways influence blood pressure and
  cardiovascular disease risk.
\newblock {\em Nature}, 478(7367):103.

\bibitem[Evans et~al., 2013]{evans2013mining}
Evans, D.~M., Brion, M. J.~A., Paternoster, L., Kemp, J.~P., McMahon, G.,
  Munaf{\`o}, M., Whitfield, J.~B., Medland, S.~E., Montgomery, G.~W., Timpson,
  N.~J., et~al. (2013).
\newblock Mining the human phenome using allelic scores that index biological
  intermediates.
\newblock {\em PLoS Genet}, 9(10):e1003919.

\bibitem[Evans and Davey~Smith, 2015]{evans2015mendelian}
Evans, D.~M. and Davey~Smith, G. (2015).
\newblock Mendelian randomization: new applications in the coming age of
  hypothesis-free causality.
\newblock {\em Annual Review of Genomics and Human Genetics}, 16:327--350.

\bibitem[Frahm, 2004]{frahm2004generalized}
Frahm, G. (2004).
\newblock {\em Generalized elliptical distributions: theory and applications}.
\newblock PhD thesis, Universit{\"a}tsbibliothek.

\bibitem[Hansen et~al., 2008]{hansen2008estimation}
Hansen, C., Hausman, J., and Newey, W. (2008).
\newblock Estimation with many instrumental variables.
\newblock {\em Journal of Business \& Economic Statistics}, 26(4):398--422.

\bibitem[Hemani et~al., 2016]{hemani2016mr}
Hemani, G., Zheng, J., Wade, K.~H., Laurin, C., Elsworth, B., Burgess, S.,
  Bowden, J., Langdon, R., Tan, V., Yarmolinsky, J., et~al. (2016).
\newblock {MR-Base}: a platform for systematic causal inference across the
  phenome using billions of genetic associations.
\newblock {\em BioRxiv}, page 078972.

\bibitem[Initiative et~al., 2020]{covid2020covid}
Initiative, C.-. H.~G. et~al. (2020).
\newblock The covid-19 host genetics initiative, a global initiative to
  elucidate the role of host genetic factors in susceptibility and severity of
  the sars-cov-2 virus pandemic.
\newblock {\em European Journal of Human Genetics}, 28(6):715.

\bibitem[Initiative et~al., 2021]{covid2021mapping}
Initiative, C.-. H.~G. et~al. (2021).
\newblock Mapping the human genetic architecture of covid-19 by worldwide
  meta-analysis.
\newblock {\em MedRxiv}.

\bibitem[Kotz and Nadarajah, 2004]{kotz2004multivariate}
Kotz, S. and Nadarajah, S. (2004).
\newblock {\em Multivariate t-distributions and their applications}.
\newblock Cambridge University Press.

\bibitem[Lambert et~al., 2013]{lambert2013meta}
Lambert, J.-C., Ibrahim-Verbaas, C.~A., Harold, D., Naj, A.~C., Sims, R.,
  Bellenguez, C., Jun, G., DeStefano, A.~L., Bis, J.~C., Beecham, G.~W., et~al.
  (2013).
\newblock Meta-analysis of 74,046 individuals identifies 11 new susceptibility
  loci for alzheimer's disease.
\newblock {\em Nature genetics}, 45(12):1452--1458.

\bibitem[Lawlor et~al., 2008]{lawlor2008mendelian}
Lawlor, D.~A., Harbord, R.~M., Sterne, J.~A., Timpson, N., and Davey~Smith, G.
  (2008).
\newblock Mendelian randomization: using genes as instruments for making causal
  inferences in epidemiology.
\newblock {\em Statistics in Medicine}, 27(8):1133--1163.

\bibitem[Liu et~al., 1998]{liu1998parameter}
Liu, C., Rubin, D.~B., and Wu, Y.~N. (1998).
\newblock Parameter expansion to accelerate em: the px-em algorithm.
\newblock {\em Biometrika}, 85(4):755--770.

\bibitem[Locke et~al., 2015]{locke2015genetic}
Locke, A.~E., Kahali, B., Berndt, S.~I., Justice, A.~E., Pers, T.~H., Day,
  F.~R., Powell, C., Vedantam, S., Buchkovich, M.~L., Yang, J., et~al. (2015).
\newblock Genetic studies of body mass index yield new insights for obesity
  biology.
\newblock {\em Nature}, 518(7538):197--206.

\bibitem[MacArthur et~al., 2017]{macarthur2017new}
MacArthur, J., Bowler, E., Cerezo, M., Gil, L., Hall, P., Hastings, E.,
  Junkins, H., McMahon, A., Milano, A., Morales, J., et~al. (2017).
\newblock The new {NHGRI-EBI} {Catalog} of published genome-wide association
  studies (gwas catalog).
\newblock {\em Nucleic Acids Research}, 45(D1):D896--D901.

\bibitem[Martens et~al., 2006]{martens2006instrumental}
Martens, E.~P., Pestman, W.~R., de~Boer, A., Belitser, S.~V., and Klungel,
  O.~H. (2006).
\newblock Instrumental variables: application and limitations.
\newblock {\em Epidemiology}, pages 260--267.

\bibitem[Pickrell et~al., 2016]{pickrell2016detection}
Pickrell, J.~K., Berisa, T., Liu, J.~Z., S{\'e}gurel, L., Tung, J.~Y., and
  Hinds, D.~A. (2016).
\newblock Detection and interpretation of shared genetic influences on 42 human
  traits.
\newblock {\em Nature Genetics}, 48(7):709.

\bibitem[Purcell et~al., 2007]{purcell2007plink}
Purcell, S., Neale, B., Todd-Brown, K., Thomas, L., Ferreira, M.~A., Bender,
  D., Maller, J., Sklar, P., De~Bakker, P.~I., Daly, M.~J., et~al. (2007).
\newblock {PLINK}: a tool set for whole-genome association and population-based
  linkage analyses.
\newblock {\em The American Journal of Human Genetics}, 81(3):559--575.

\bibitem[Rothman, 2012]{rothman2012positive}
Rothman, A.~J. (2012).
\newblock Positive definite estimators of large covariance matrices.
\newblock {\em Biometrika}, 99(3):733--740.

\bibitem[Schunkert et~al., 2011]{schunkert2011large}
Schunkert, H., K{\"o}nig, I.~R., Kathiresan, S., Reilly, M.~P., Assimes, T.~L.,
  Holm, H., Preuss, M., Stewart, A.~F., Barbalic, M., Gieger, C., et~al.
  (2011).
\newblock Large-scale association analysis identifies 13 new susceptibility
  loci for coronary artery disease.
\newblock {\em Nature Genetics}, 43(4):333--338.

\bibitem[Solovieff et~al., 2013]{solovieff2013pleiotropy}
Solovieff, N., Cotsapas, C., Lee, P.~H., Purcell, S.~M., and Smoller, J.~W.
  (2013).
\newblock Pleiotropy in complex traits: challenges and strategies.
\newblock {\em Nature Reviews Genetics}, 14(7):483--495.

\bibitem[Teslovich et~al., 2010]{teslovich2010biological}
Teslovich, T.~M., Musunuru, K., Smith, A.~V., Edmondson, A.~C., Stylianou,
  I.~M., Koseki, M., Pirruccello, J.~P., Ripatti, S., Chasman, D.~I., Willer,
  C.~J., et~al. (2010).
\newblock Biological, clinical and population relevance of 95 loci for blood
  lipids.
\newblock {\em Nature}, 466(7307):707--713.

\bibitem[Van~der Vaart, 2000]{van2000asymptotic}
Van~der Vaart, A.~W. (2000).
\newblock {\em Asymptotic Statistics}, volume~3.
\newblock Cambridge University Press.

\bibitem[Verbanck et~al., 2018]{verbanck2018detection}
Verbanck, M., Chen, C.-y., Neale, B., and Do, R. (2018).
\newblock Detection of widespread horizontal pleiotropy in causal relationships
  inferred from mendelian randomization between complex traits and diseases.
\newblock {\em Nature Genetics}, 50(5):693--698.

\bibitem[Wang and Titterington, 2005]{wang2005inadequacy}
Wang, B. and Titterington, D. (2005).
\newblock Inadequacy of interval estimates corresponding to variational
  bayesian approximations.
\newblock In {\em AISTATS}. Citeseer.

\bibitem[Willer et~al., 2013]{willer2013discovery}
Willer, C.~J., Schmidt, E.~M., Sengupta, S., Peloso, G.~M., Gustafsson, S.,
  Kanoni, S., Ganna, A., Chen, J., Buchkovich, M.~L., Mora, S., et~al. (2013).
\newblock Discovery and refinement of loci associated with lipid levels.
\newblock {\em Nature Genetics}, 45(11):1274.

\bibitem[Yang et~al., 2018]{yang2018lpg}
Yang, Y., Dai, M., Huang, J., Lin, X., Yang, C., Chen, M., and Liu, J. (2018).
\newblock {LPG}: A four-group probabilistic approach to leveraging pleiotropy
  in genome-wide association studies.
\newblock {\em BMC Genomics}, 19(1):503.

\bibitem[Yang et~al., 2020]{yang2020comm}
Yang, Y., Shi, X., Jiao, Y., Huang, J., Chen, M., Zhou, X., Sun, L., Lin, X.,
  Yang, C., and Liu, J. (2020).
\newblock {CoMM-S2}: a collaborative mixed model using summary statistics in
  transcriptome-wide association studies.
\newblock {\em Bioinformatics}, 36(7):2009--2016.

\bibitem[Zhao et~al., 2020a]{zhao2020bayesian}
Zhao, J., Ming, J., Hu, X., Chen, G., Liu, J., and Yang, C. (2020a).
\newblock Bayesian weighted mendelian randomization for causal inference based
  on summary statistics.
\newblock {\em Bioinformatics}, 36(5):1501--1508.

\bibitem[Zhao et~al., 2020b]{zhao2020statistical}
Zhao, Q., Wang, J., Hemani, G., Bowden, J., and Small, D.~S. (2020b).
\newblock Statistical inference in two-sample summary-data mendelian
  randomization using robust adjusted profile score.
\newblock {\em Annals of Statistics}, 48(3):1742--1769.

\bibitem[Zhu and Stephens, 2017]{zhu2017bayesian}
Zhu, X. and Stephens, M. (2017).
\newblock Bayesian large-scale multiple regression with summary statistics from
  genome-wide association studies.
\newblock {\em The Annals of Applied Statistics}, 11(3):1561.

\bibitem[Zhu et~al., 2018]{zhu2018causal}
Zhu, Z., Zheng, Z., Zhang, F., Wu, Y., Trzaskowski, M., Maier, R., Robinson,
  M.~R., McGrath, J.~J., Visscher, P.~M., Wray, N.~R., et~al. (2018).
\newblock Causal associations between risk factors and common diseases inferred
  from gwas summary data.
\newblock {\em Nature Communications}, 9(1):1--12.

\end{thebibliography}

\end{document}